\documentclass[11pt, a4paper]{article}
\usepackage[english]{babel}
\usepackage{times}
\usepackage[centertags,reqno]{amsmath}
\usepackage{amssymb}
\usepackage{jpart}

\nc\Input{\input}
\nc\betah{\hat \beta}
\nc\ubl{\overline{\lambda}}
\nc{\laa}{\mathcal A}
\nc{\gie}{\mathcal G}
\nc\ha{\mathcal{H}}
\nc\gk{\mathcal G}
\nc\ga{\mathcal A}
\nc{\prob}{\mathbb{P}}
\nc\pe{\mathcal{P}}
\nc{\lde}{\mathcal{D}}
\nc\eh{\mathcal{H}}
\nc{\ef}{\mathcal{F}}
\nc{\er}{\mathbb{R}}
\nc{\erp}{\mathbb{R}_+}
\nc{\en}{{\mathbb{N}}}
\nc{\es}{\mathcal{S}}
\nc{\te}{\mathcal{T}}
\nc{\borel}{\mathcal{B}}
\nc\bse{\mathcal{E}}
\nc\flz{\mathcal{Z}}
\nc\flm{\mathcal{M}}
\nc\fly{\mathcal{Y}}
\nc{\tl}{\tilde}
\nc{\ee}{{\mathbb{E}\,}}
\nc{\cee}[3]{{\ee #3[#1\,|\,#2 #3]}}
\nc{\norma}[1]{{\| #1 \|}}
\nc{\ind}[1]{1_{#1}}
\nc{\ilskal}[2]{{\langle {#1} , {#2} \rangle}}
\nc{\setcond}{: \exs}
\nc{\supp}{\mathop{\rm supp} \nolimits}
\nc{\conv}{\mathop{\rm conv} \nolimits}
\nc{\essinf}{\mathop{\rm ess\, inf} \limits}
\nc{\esssup}{\mathop{\rm ess\, sup} \limits}
\nc{\vvect}[2]{{\begin{pmatrix} #1 \\ #2 \end{pmatrix}}}
\nc{\vvvect}[3]{{\begin{pmatrix} #1 \\ #2 \\ #3 \end{pmatrix}}}
\nc\cont[1]{C^{\mathcal #1}}

\nc{\costa}{c\big(N_{i-1}, N_i, S(\tau_i)\big)}
\nc{\costb}[1][]{c\big(\tl N_{i-1}, \tl N_i #1, S(\tau_i)\big)}
\nc\devnull[1]{}

\begin{document}


\author{
Jan Palczewski\footnote{
School of Mathematics, University of Leeds, Leeds LS2 9JT, UK
and
Faculty of Mathematics, University of Warsaw, Banacha 2, 02-097 Warszawa, Poland
(e-mail: J.Palczewski@mimuw.edu.pl)}
\and
\L ukasz Stettner\footnote{
Institute of Mathematics, Polish Academy of Sciences, Sniadeckich 8, 00-950 Warszawa, Poland,
(e-mail: \mbox{stettner@impan.gov.pl}).}}

\title{Growth-optimal portfolios under transaction costs}
\maketitle

\begin{abstract}
This paper studies a portfolio optimization problem in a discrete-time Markovian model of a financial market, in which asset price dynamics depend on an external process of economic factors. There are transaction costs with a structure that covers, in particular, the case of fixed plus proportional costs. We prove that there exists a self-financing trading strategy maximizing the average growth rate of the portfolio wealth. We show that this strategy has a Markovian form. Our result is obtained by large deviations estimates on empirical measures of the price process and by a generalization of the vanishing discount method to discontinuous transition operators.
\end{abstract}

\n \textbf{Keywords: } portfolio optimization, transaction costs, growth rate, logarithmic utility, Markov process, impulsive strategy, vanishing discount

\setcounter{secnumdepth}{2}

\section{Introduction}
Researchers and practitioners have long been aware that Markovian models of asset price dynamics, such as the Cox-Ross-Rubinstein model or the Black-Scholes model, have significant deficiencies related to non-stationarity of the financial market. They observed that the volatility and the expected rate of return of asset prices are not constant but depend on an economic situation, which may change over longer time spans. As a remedy, they  introduced additional processes modeling vital market variables, such as market trend or price volatility. However, a unified  framework has only recenly been introduced and has attracted a lot of interest (see eg. \cite{Bielecki2}, \cite{Bielecki1}, \cite{fleming}, \cite{Palczewski1}, \cite{Palczewski2}, \cite{Stettner2}). Existing literature concentrates mainly on continuous-time diffusion models. Bielecki et al. \cite{Bielecki1} solve an asset management problem where economic factors, as those additional market variables are called, form a diffusion that is independent of the Brownian motion governing the price process and they affect only the drift of the price process. Fleming and Sheu \cite{fleming} allow both processes to have dependent Brownian motions but their diffusions are of a special form. Palczewski and Stettner \cite{Palczewski1}, though, assume only that asset prices and economic factors follow one general continuous-time Markov process and prove results concerning optimal portfolio selection for infinite time disounted performance functional under transaction costs. 

In the present paper we study a portfolio management problem in which performance is measured by an average growth rate of the portfolio wealth. We work within a discrete time framework which allows us to overcome limitations and technicalities of the existing theory of continous time Markov processes and impulsive control.  The market consists of $d$ assets, whose prices are, in general, interdependent. Their dynamics are affected by a process of economic factors, which is a Markov process on a Polish space (for details see Section \ref{sec:a}). We assume that assets cannot go bankrupt (their prices are positive). We impose costs of performing transactions. These costs, in the simplest, consist of a fixed part, independent of the transaction, and a proportional part, depending on the volume and the type of assets sold or purchased (see (\ref{eqn:a01}), (\ref{eqn:a01a}) and the following discussion). This type of transaction costs prevents continuous trading in continuous-time models (see e.g. \cite{Palczewski1}) and emulates existing market mechanisms.  The framework of this paper covers more general transaction costs structures as well (see Section \ref{sec:g}).
Performance of a portfolio $\Pi$ is measured by the funtional
\begin{equation}
 \label{eqn:001}
J(\Pi) = \liminf_{T \to \infty} \frac{1}{T} \ee \ln X^\Pi(T),
\end{equation}
where $X^\Pi(T)$ is the wealth of the portfolio $\Pi$ at time $T$. This functional computes an average growth rate of the portfolio $\Pi$ as can be seen from the following reformulation of the above formula:
\begin{equation}
 \label{eqn:0011}
J(\Pi) = \liminf_{T \to \infty} \frac{1}{T} \ee \sum_{k=0}^{T-1}
\ln \frac{X^\Pi(k+1)}{X^\Pi(k)}.
\end{equation}
The aim of this paper is to find a portfolio that maximizes the value of (\ref{eqn:001}). 
This is  an infinite-time counterpart of the logarithmic utility maximization, which is widely used in the economic and financial community, where optimal portfolios are refered to as log-optimal or growth-optimal. For a broader treatment see textbooks \cite{duffie}, \cite{luenberger}. In mathematical context the research goes back to Kelly (see \cite{kelly}, \cite{thorpe}) and has continued in discrete time (\cite{algoet}) and continuous time (\cite{aase}, \cite{Akian}) up to today (\cite{gerenc}, \cite{inoue}, \cite{Platen}). Functional (\ref{eqn:001}) can also be seen as a risk sensitive functional and the literature is here broad as well (\cite{Bielecki1}, \cite{kuroda}, \cite{Stettner2}). It should be stressed that the majority of papers considers continuous time diffusion models, where an optimal strategy is obtained as a solution to an appropriate HJB equation, usually reformulated in a variational form. Consequently, the results are based on a sophisticated theory of PDE's and solutions usually do not use directly probabilistic properties of the phenomena under study. Moreover, due to complexity of the studied PDEs the results are often of existential form. 

In this paper, we approach the optimization problem (\ref{eqn:001}) from a probabilistic point of view. We prove that there exists a self-financing portfolio strategy maximizing the growth-rate (\ref{eqn:001}). We show that this trading strategy has a Markovian form, i.e. an investment decision at time $t$ is based only on the state of asset prices and economic factors at $t$. Main additions to the existing theory are transaction costs with a fixed term and a general form of dependence of asset prices on economic factors. As far as we know there is no paper that treats this type of problems in such generality.

Our study depends strongly on the reformulation (\ref{eqn:0011}) of the performace functional. It exposes the Markovian structure of the functional and allows application of the theory of optimization of long-run average cost functionals. A survey of standard methods for long-run average cost functionals is in \cite{Arapost}. We will, however, borrow from a new technique invented by Sch\"al \cite{Schal}, who initiated use of Bellman inequalities leading to significantly more general results. His ideas thrive in \cite{Hernandez1} (weighted norms), \cite{gonzalez} (stochastic games) and recently in \cite{jaskiewicz}. Those results strongly depend on continuity properties of the controlled transition operator of the Markov process under consideration. In this paper we show that the above ideas can also be used in the study of problems which violate the continuity assumptions. Moreover, following \cite{jaskiewicz} we are able to remove a requirement for the state space to be locally compact as is needed in the seminal paper \cite{Schal}. This significantly generalized the applicability of this framework to incomplete information case (for details on the incomplete information model see Section \ref{sec:g} and \cite{Palczewski2}).

The paper is organized as follows. In Section \ref{sec:a} we introduce the model. We specify the dynamics of asset price process and the form of transaction costs. We introduce a process representing proportions of the portfolio wealth invested in the individual assets and we reformulate the initial problem in terms of proportions. This reformulation plays a major role in the paper.

Section \ref{sec:b} sees main assumptions presented. We prove ergodicity results and large deviation estimates on empirical measures of the price process. They give a new insight into the dynamics of the price process.

The study of value functions of discounted functionals related to (\ref{eqn:0011}) is pursued in Section \ref{sec:c}. A few important technical results on the consequences of the transaction costs are stated (their proofs are in Appendix). They are used to build a relation between value functions for the discounted problems with and without the fixed term in the transaction costs structure. It is a starting point for derivation of the Bellman inequality, which is performed in Section \ref{sec:d}. At this stage we also deal with the lack of continuity of the controlled transition operator. We prove the existence of a growth-optimal strategy and show its form. We also relate the results to the case without constant term in the transaction costs structure (see \cite{Stettner2}).

Section \ref{sec:g} presents extensions of our results to other transaction costs structures and shows how existing results can be used in the case of incomplete observation of economic factors.

\section{Preliminaries}
\label{sec:a}

The market model is constructed on a probability space $(\Omega, \ef, \prob)$. Prices of $d$ assets are represented by the process
$\big(S(t)\big)_{t = 0, 1, \ldots}$, 
$S(t) = \big(S^1(t), \ldots, S^d(t)\big) \in (0, \infty)^d$. Economic factors are modeled by a time homogeneous Markov process $\big(Z(t)\big)_{t = 0, 1, \ldots}$ with values in a Polish (separable, complete, metric) space $E$ with Borel $\sigma$-algebra $\bse$.  The dynamics of the price process are governed by the equation
\begin{equation}
\label{eqn:a07}
\frac{S^i(t+1)}{S^i(t)} = \zeta^i\Big(Z(t+1), \xi(t+1)\Big),
\frs S^i(0) = s^i > 0, \frs i = 1, \ldots, d,
\end{equation}
where $\big(\xi(t)\big)_{t=1, 2, \ldots}$ is a sequence of i.i.d. random variables
with values in a Polish space $(E^\xi, \bse^\xi)$ and functions $\zeta^i: (E, \bse) \times (E^\xi, \bse^\xi) \to (0, \infty)$ are Borel measurable, $i = 1, \ldots, d$.  We assume that $\big(S(t), Z(t)\big)$ forms a weak Feller process, i.e. its transition operator transforms the space of bounded continuous functions into itself.
In the sequel we shall write $\zeta^i(t)$ for $\zeta^i\Big(Z(t), \xi(t)\Big)$, and
$\zeta(t)$ for the vector $\big(\zeta^1(t), \ldots, \zeta^d(t)\big)$, whenever it does not lead to ambiguity. 

Let $\big(\ef_t\big)$ be a filtration generated by $\big(S(t), Z(t)\big)$. It represents the knowledge of an investor observing the market. Therefore, in general, it depends on the starting point of the process  $\big(S(t), Z(t)\big)$. Due to (\ref{eqn:a07}) the filtration is in fact independent of $S(0)$. It depends only on the initial value $z$ of the process $\big(Z(t)\big)$, and to stress this dependence it will be denoted by $\big(\ef^z_t\big)$.

Fix $s \in (0, \infty)^d$ and $z \in E$, the initial values of processes $\big(S(t)\big)$ and $\big(Z(t)\big)$.
A trading strategy is a sequence of pairs $\big((N_k, \tau_k)\big)_{k=0, 1,\ldots}$, where
$\tau_0 = 0$, $(\tau_k)_{k=1,2, \ldots}$ are $\big(\ef^z_t\big)$ stopping times, and $\tau_{k+1} > \tau_k$, $k=1, 2, \ldots$. Stopping times $(\tau_k)$, $k \ge 1$, represent moments of transactions, whereas $\tau_0=0$ is only introduced for convenience of notation. The number of shares held in the portfolio in the time interval $[\tau_k, \tau_{k+1})$ is denoted by $N_k$, which is an $\ef^z_{\tau_k}$-measurable random variable with values in $[0, \infty)^d$. Hence, $N_0$ is a deterministic initial portfolio.

The share holding process is given by
$$
N(t) = \sum_{k=1}^\infty \ind{t \in [\tau_k, \tau_{k+1})} N_k, \quad t \ge 0.
$$
In what follows we shall consider transaction costs of one of the forms
\begin{align}
\label{eqn:a01}
\tl c(\eta_1, \eta_2, S) &= \sum_{i=1}^d \Big( c^1_i S^i (\eta_1^i - \eta_2^i)^+ + c^2_i S^i (\eta_1^i - \eta_2^i)^-
\big) + C,\\
\label{eqn:a01a}
\tl c(\eta_1, \eta_2, S) &= \max\bigg(C,  \sum_{i=1}^d \Big( c^1_i S^i (\eta_1^i - \eta_2^i)^+ + c^2_i S^i (\eta_1^i - \eta_2^i)^-
\big)\bigg)
\end{align}
where $c^1_i, c^2_i \in [0,1) $ are proportional costs, $C \ge 0$, $S$ stands for the asset prices at the moment of transaction, $\eta_1$ denotes the portfolio contents before transaction, and $\eta_2$ -- after transaction.
We impose a self-financing condition on portfolios, i.e.
\begin{equation}
\label{eqn:a02}
N_{k} \cdot S(\tau_k) = N_{k-1} \cdot S(\tau_k) + \tl c\big( N_{k-1}, N_k, S(\tau_k) \big), \exs k = 1, 2, \ldots.
\end{equation}
Notice that due to the lower bound $C$ on the transaction costs function, transactions cannot be executed if the wealth
of the portfolio is smaller than $C$. It is also clear that (\ref{eqn:a02}) depends on the initial value $(s,z)$ of the process
$\big(S(t), Z(t)\big)$.

For the clarity of presentation, we shall restrict our attention to the costs of the form (\ref{eqn:a01}). However, all the results are easily modified to fit (\ref{eqn:a01a}), and, in fact, they extend to a larger family of transaction costs structures, see Section \ref{sec:g}.

In the case of no transaction costs or proportional transactions costs it is natural to reformulate the problem in terms of proportions as it transforms the set of controls (possible portfolios) to a compact set, which faciliates matematical analysis. In our more general framework, we shall also benefit from this reformulation.

Denote by $X_-(t)$ the wealth of the portfolio before a possible transaction at $t$ and by $X(t)$ the wealth just after the transaction:
\begin{equation}
\begin{aligned}
&X(t) = N(t) \cdot S(t), \\
&X_-(t) = N(t-1) \cdot S(t).
\end{aligned}
\end{equation}
If there is no transaction at $t$ both values are identical.
In a similar way, for $i=1, \ldots, d$, we construct two processes representing proportions of our capital invested in the asset $i$:
\begin{equation}
\begin{aligned}
&\pi^i(t) = \frac{N^i(t) S^i(t)}{X(t)},\\
&\pi_-^i(t) = \frac{N^i(t-1) S^i(t)}{X_-(t)}.
\end{aligned}
\end{equation}
Since short sales are prohibited we have $\pi(t), \pi_-(t) \in \es$, where $\es$ is the unit simplex in $\er^d$:
$$
\es = \{ (\pi^1, \ldots, \pi^d) \setcond \pi^i \ge 0, \exs \sum_{i=1}^d \pi^i = 1 \}.
$$
Denote by $\es^0$ the polyhedral set generated by $\es$:
$$
\es^0 = \{ (\pi^1, \ldots, \pi^d) \setcond \pi^i \ge 0, \exs \sum_{i=1}^d \pi^i \le 1 \}
$$
and let $g: \es^0 \to \es$ be a projection from $\es^0$ to its boundary $\es$
$$
g(\pi^1, \ldots, \pi^d) = \Big(\frac{\pi^1}{\sum \pi^i}, \ldots, \frac{\pi^d}{\sum \pi^i}\Big).
$$
Define
$$
c(\pi_-, \tl \pi) = \sum_{i=1}^d \Big( c^1_i (\tl \pi^i - \pi_-^i)^+ + c^2_i (\tl \pi^i - \pi_-^i)^-\big).
$$
The self-financing condition (\ref{eqn:a02}) can be written as
\begin{equation}
\label{eqn:a03}
X_-(\tau_k) = X(\tau_k) + X_-(\tau_k) \Big( c\big(\pi_-(\tau_k), \tl \pi_k\big) + \frac{C}{X_-(\tau_k)} \Big), \exs k = 1,2, \ldots
\end{equation}
for some $\tl \pi_k \in \es^0$ such that  $\pi(\tau_k) = g\big(\tl \pi_k \big)$. From (\ref{eqn:a02}) one can deduce that $\tl \pi_k = \frac{X(\tau_k)}{X_-(\tau_k)} \pi(\tau_k)$ satisfies (\ref{eqn:a03}).

Given $\pi_-, \pi \in \es$, $x_- > 0$ define a function
$$
F^{\pi_-, \pi, x_-}(\delta) =
c\big(\pi_-, \delta \pi\big) + \frac{C}{x_-} + \delta.
$$
Equation (\ref{eqn:a03}) can be written equivalently as
$$
F^{\pi_-(\tau_k), \pi(\tau_k), X_-(\tau_k)} \Big(\sum_{i=1}^d \tl \pi^i_k\Big) = 1.
$$
The following lemma states a crucial property of $F$ that will be used to reformulate the self-financing condition.
\bl
\label{lem:a01}
There exists a unique function
$e: \es \times \es \times (0, \infty) \to [0,1]$, such that
\begin{itemize}
\item[(1)] if $e(\pi_-, \pi, x) > 0$, then $F^{\pi_-, \pi, x_-}\big(e(\pi_-, \pi, x_-)\big) = 1$,
\item[(2)] $e(\pi_-, \pi, x_-) = 0$ if and only if the equation $F^{\pi_-, \pi, x_-}(\cdot) = 1$ has no solution in $(0,1]$.
\end{itemize}
Moreover, $e$ is continuous.
\el
\dowod
The proof is rather straightforward and resembles the proof of Lemma 1 in \cite{Stettner2}.
\qed

Let $\big((N_k, \tau_k)\big)$ be a self-financing trading strategy and let $\pi_-(\tau_k)$, $\pi(\tau_k)$ be defined as above.
By virtue of Lemma \ref{lem:a01} for any $k\in \en$ we have
$$
F^{\pi_-(\tau_k), \pi(\tau_k), X_-(\tau_k)} \Big(e\big( \pi_-(\tau_k), \pi(\tau_k), X_-(\tau_k) \big)\Big) = 1
$$
and $\frac{X(\tau_k)}{X_-(\tau_k)} = e\big( \pi_-(\tau_k), \pi(\tau_k), X_-(\tau_k) \big)$. Second assertion is a consequence of
the uniqueness of $e$ and equation (\ref{eqn:a03}).  Therefore, any transaction can be described solely by means of proportions
$\pi_-(\tau_k)$ and $\pi(\tau_k)$, and the portfolio wealth $X_-(\tau_k)$. Consequently, any trading strategy has a unique representation in the following form: the initial wealth $x_- = N_0 \cdot S(0)$, the initial proportion
$$
\pi_- = \bigg( \frac{N^1_0 S^1(0)}{N_0 \cdot S(0)}, \ldots, \frac{N^d_0 S^d(0)}{N_0 \cdot S(0)}\bigg),
$$
and $\big((\pi_k, \tau_k)\big)_{k=1, 2, \cdots}$, where $\pi_k$ is the post-transaction proportion represented by an $\es$-valued
$\ef_{\tau_k}$-measurable random variable. Indeed, define the corresponding pre-transaction proportion process $\pi_-(t)$ by
\begin{equation}
\label{eqn:a09}
\begin{aligned}
&\pi_-(0) = \pi_-,\\
&\pi_-(t) = \pi_{k} \diamond \zeta(\tau_k+1) \diamond \ldots \diamond \zeta(t),
\frs \tau_{k} < t \le \tau_{k+1},
\end{aligned}
\end{equation}
where for simplicity of the notation we have $\tau_0 = 0$ and
\begin{equation}
\label{eqn:a10}
\pi \diamond \zeta = g(\pi^1 \zeta^1, \ldots, \pi^d \zeta^d),
\frs \pi \in \es,\exs \zeta \in (0, \infty)^d.
\end{equation}
The corresponding post-transaction proportion process is given by
\begin{equation}
\label{eqn:a11}
\begin{aligned}
&\pi(t) = \begin{cases}
\pi_-, & t = 0 \text{ and } \tau_1 > 0\\
\pi_{k}, & t = \tau_k\\
\pi_{k} \diamond \zeta(\tau_k+1) \diamond \ldots \diamond \zeta(t),& \tau_{k} < t < \tau_{k+1}
          \end{cases}
.\end{aligned}
\end{equation}
At the moment $\tau_k$ the pre-transaction wealth $X_-(\tau_k)$ is diminished to
$$
X(\tau_k) = X_-(\tau_k) e\big(\pi_-(\tau_k), \pi(\tau_k), X_-(\tau_k)\big).
$$
Furthermore,
$$
X_-(t+1) = \sum_{i=1}^d \frac{\pi^i(t) X(t)}{S^i(t)} S^i(t+1)
= X(t) \Big( \pi(t) \cdot \zeta(t+1) \Big).
$$
Consequently,
\begin{equation}
\label{eqn:a04}
X_-(t) = X_-(0) \prod_{s=0}^{t-1} \Big(\pi(s) \cdot \zeta(s + 1)\Big)
\prod_{k=1}^\infty \Big(
\ind{\tau_k < t} e\big(\pi_-(\tau_k), \pi(\tau_k), X_-(\tau_k)\big)
+ \ind{\tau_k \ge t} \Big),
\end{equation}
which finishes the construction of the correspondence between the primal definition of a trading strategy with the share holding process $N(t)$ and the equivalent form with proportions.
Notice that due to our reformulation, the self-financing condition no longer depends on the initial value of the asset price process $\big(S(t)\big)$.

Let $\ga^z$ be a set of sequences $\big((\pi_k, \tau_k)\big)_{k=1,2,\ldots}$, where $\tau_k$ is an $(\ef^z_t)$ stopping time, $\pi_k$ is an $\ef^z_{\tau_k}$-measurable random variable with values in $\es$, and $\tau_{k+1} > \tau_k$, $k=1, 2, \ldots$. Elements of $\ga^z$ will be called admissible trading strategies or admissible portfolios. Notice that for a fixed initial wealth $x_-$ and an initial proportion $\pi_-$  not every admissible trading strategy $\big((\pi_k, \tau_k)\big)$ is related to some self-financing strategy $\big((N_k, \tau_k)\big)$. Indeed, if $X_-(\tau_k)$ is small for some $k$, not all proportions $\pi(\tau_k)$ are attainable from $\pi_-(\tau_k)$. If $\pi(\tau_k)$ is attainable, we have
$$
F^{\pi_-(\tau_k), \pi(\tau_k), X_-(\tau_k)} \Big(e\big(\pi_-(\tau_k), \pi(\tau_k), X_-(\tau_k)\big)\Big) = 1
$$
and
$$
X(\tau_k) = X_-(\tau_k) e\big(\pi_-(\tau_k), \pi(\tau_k), X_-(\tau_k)\big).
$$
If $\pi(\tau_k)$ is not attainable, we have, by the above construction,
$$
e\big(\pi_-(\tau_k), \pi(\tau_k), X_-(\tau_k)\big) = 0
$$
and
$$
X(\tau_k) = 0.
$$
Therefore, in what follows we may assume that all proportions are attainable from $\pi_-(\tau_k)$ irrespective of the value of $X_-(\tau_k)$, but they may lead to zero wealth process if we cannot afford to pay transaction costs. Since the strategy allowing anihilation of wealth is not optimal (the functional in (\ref{eqn:0011}) evaluates to $-\infty$), the extension of the set of trading strategies does not have any impact on optimal strategies.

As we noticed before, the set of admissible strategies and the wealth of the portfolio are independent of the initial prices of the assets. Therefore, instead of writing $\prob^{(s,z)}$ and $\ee^{(s,z)}$ to stress dependence of the probability measure on the initial state of the Markov process $\big(S(t), Z(t)\big)$ we will write $\prob^z$ and $\ee^z$.

The goal of this paper is to maximize the functional
\begin{equation}
\label{eqn:a05}
J^{\pi_-, x_-, z} (\Pi) = \liminf_{T \to \infty} \frac{1}{T} \ee^z \ln X_-(T)
\end{equation}
over all portfolios $\Pi \in \ga^z$, where $\pi_-$ is the initial proportion, $x_-$ denotes the initial wealth and $z$ is the initial state of the economic factor process. Observe that using (\ref{eqn:a04}) we obtain
\begin{equation}
\label{eqn:a06}
\begin{aligned}
J^{\pi_-, x_-, z} (\Pi) = \liminf_{T \to \infty} \frac{1}{T}
\bigg\{ &\sum_{t=0}^{T - 1} \ee^z \ln \pi(t) \cdot \zeta(t + 1) \\
&+
\sum_{k=1}^\infty \ee^z \Big\{ \ind{\tau_{k} < T} \ln e\big(\pi_-(\tau_{k}), \pi_k, X_-(\tau_{k})\big) \Big\} \bigg\}.
\end{aligned}
\end{equation}
This transformes our problem to the form suitable for further analysis.

\section{Assumptions and basic properties of the price process}
\label{sec:b}
Denote by $P(z, dy)$ the transition operator of the process $\big(Z(t)\big)$. Let $\hat E = E \times E^\xi$, and let $\nu$ be the law of $\xi(1)$ on $E^\xi$. For $x = (z, \xi)$ and a bounded measurable function $w$ on $\hat E$ define
$$
\hat Pw(x) = \int_E \int_{E^\xi} w(z', \xi') \nu(d\xi') P(z, dz').
$$
Consider the following assumptions:
\begin{opisanie}{(A1)}
The process $\big(S(t), Z(t)\big)$ satisfies the Feller property i.e. its transition operator maps the space of continuous bounded functions into itself.
\end{opisanie}
\begin{opisanie}{(A2)}
$\displaystyle \es \times E \ni (\pi, z) \mapsto h(\pi, z) =
\ee^z \big\{ \ln \pi \cdot \zeta\big(Z(1), \xi(1)\big) \big\}$ is a bounded, continuous function.
\end{opisanie}
\begin{opisanie}{(A3)}
$\displaystyle \sup_{z, z' \in E} \sup_{B \in \bse} \big( P^n(z, B) - P^n(z', B) \big) = \kappa < 1$ for some $n \ge 1$.
\end{opisanie}
\begin{opisanie}{(A4)}
There is a continuous function $\hat u_0$ defined on $\hat E$ such that $\hat u_0(x) \ge 1$ for $x \in \hat E$, the function $x \mapsto \hat P \hat u_0(x)$ is bounded on compact subsets of $\hat E$ and for any positive real number $l$ the set $\Big\{ x: \frac{\hat u_0(x)}{\hat P \hat u_0(x)} \le l \Big\}$ is compact.
\end{opisanie}
\begin{opisanie}{(A5)}
The function $\zeta(z, \xi)$ is continuous and separated from $0$, i.e.
$\inf_{z, \xi} \zeta^i(z, \xi) > 0$ for $i=1, \ldots, d$.
\end{opisanie}

Due to assumption (A3) the process $Z(t)$ is uniformly ergodic. Together with (A4)-(A5) it gives important estimates on the behaviour of the asset prices, as can be seen in the following theorem:
\bt
\label{lem:b02}
Under (A1)-(A5):
\begin{itemize}
\item[i)] the process $Z(t)$ has a unique invariant probability measure $\vartheta$.
\item[ii)] for each non-negative measurable function $f$ such that $\int_E f(z) \vartheta(dz) < \infty$
$$
\lim_{T \to \infty} \frac{1}{T} \sum_{t = 0}^T \ee^z f\big(Z(t)\big) =  \int_E f(z) \vartheta(dz).
$$
\item[iii)] The following large deviations estimate holds: for each $\epsilon > 0$ there exists $T^* > 0$, $\gamma > 0$, $K > 0$ such that for all $T \ge T^*$
$$
\prob^z \Big\{ \frac{1}{T} \ln\Big(\prod_{t=0}^{T-1} \hat \zeta\big(Z(t+1), \xi(t+1)\big)\Big) \le \hat p - \epsilon \Big\} \le
K e^{-\gamma T},
$$
where
$$
\hat \zeta(z, \xi) = \min\big( \zeta^1(z, \xi), \ldots, \zeta^d(z, \xi)\big)
$$
and
$$
\hat p = \int_{E\times E^\xi} \ln \hat\zeta(z, \xi)\, \vartheta(dz) \nu(d\xi).
$$
\end{itemize}
\et
\dowod
Notice that (A3) implies that for arbitrary $z, z' \in E$ and $B \in \bse$
$$
P^n(z, B) \le \kappa + P^n(z', B) \big).
$$
Therefore, Condition (D) (a version of Doeblin's hypothesis) in \cite{Doob}, Section V.5, holds with $\phi(B) = P^n(z', B)$ for some $z' \in E$ and $\epsilon = \frac{1-\kappa}{2}$. Applying (A3) one also gets that for any bounded measurable function $f$ we have
$$
\ee^{z_1} f\big(Z(n)\big) - \ee^{z_2} f\big(Z(n)\big) =
\int_E f(z) \big( P^n(z_1, dz) - P^n(z_2, dz)\big) \le \kappa \|f\|_\infty.
$$
Therefore, due to Theorems V.5.7 and V.6.2 in \cite{Doob} there is a unique invariant probability measure for $\big(Z(t)\big)$ and (ii) holds.

Statement (iii) results from application of the Large Deviations Theory to the Markov process $\big(Z(t), \xi(t)\big)$. Recall that the transition operator of this process is denoted by $\hat P$. A measure $\vartheta \otimes \nu$ is a unique probabilistic invariant measure of $\hat P$. Let
$$
L_n = \frac{1}{n} \sum_{t=1}^n \delta_{(Z(t),\, \xi(t))}, \qquad n=1, 2, \ldots,
$$
denote the empirical distribution of the process $\big(Z(t), \xi(t)\big)$. Notice that $L_n$ takes values in the space $\pe = \pe(E \times E^\xi)$ of probability measures on $E\times E^\xi$ with the weak convergence topology. Due to Section 4 of \cite{donsker} (see also \cite{duncan} and \cite{liptser}) there exists a convex lower semicontinuous function $J:\pe \to \er$ (called a good rate function) such that
for any compact set $\Gamma \in \borel(\pe)$ we have
$$
\limsup_{n \to \infty} \frac{1}{n} \log\bigg( \sup_{z \in E}\ \prob^z\Big(\{ \omega:
L_n(\omega) \in \Gamma \} \Big) \bigg) \le - \inf_{\mu \in \Gamma} J(\mu).
$$
Under assumption (A4) the above inequality holds for any closed set $\Gamma$ (not necessarily compact).
By Lemma 4.2 of \cite{donsker}, the set of measures $\mu \in \pe$ such that $J(\mu) \le l$ is compact for each $l \in \er$. Consequently, for a closed set $\Gamma \subset \pe$ such that $\vartheta \otimes \nu \notin \Gamma$ we have (see Proposition 1 of \cite{duncan})
$\inf_{\mu \in \Gamma} J(\mu) > 0$.

Define
$$
\hat \Gamma = \Big\{ \mu \in \pe: \int_{E\times E^\xi} \ln \hat \zeta(z, \xi)\ \mu(dz \times d\xi) \le \hat p - \epsilon \Big\}.
$$
To complete the proof it is enough to show that $\inf_{\mu \in \hat \Gamma} J(\mu) > 0$.
Due to unboundedness of $\hat \zeta$ the set $\hat \Gamma$ may not be closed in $\pe$. However, under (A5) for every $N > 0$
$$
\Gamma_N = \Big\{ \mu \in \pe: \int_{E\times E^\xi} \min\big(\ln \hat \zeta(z, \xi), N\big) \ \mu(dz \times d\xi) \le \hat p - \epsilon \Big\}
$$
is closed and $\hat \Gamma \subseteq \Gamma_N$. Due to monotone convergence theorem there exists $N$ such that $\vartheta \otimes \nu \notin \Gamma_N$, and consequently
$\inf_{\mu \in \hat \Gamma_N} J(\mu) > 0$.
\qed

Statement (iii) of the above lemma reads that whenever the average one-step growth rate of the asset prices $\hat p$ is positive then  the prices grow exponentially fast on a large subset of $\Omega$, i.e. for $T > T^*$
$$
\prob \Big\{ S^i(T) \ge S^i(0) e^{T(\hat p - \epsilon)} \ \  \forall\, i=1, \ldots d \Big\} \ge
1 - K e^{-\gamma T}.
$$
This is a surprising result, since the condition $\hat p > 0$ is often viewed as a prerequisite for investors to be willing to invest on the market. Therefore, promising markets offer exponential speed of growth of investors' wealth.

The following remarks explain the assumptions (A1)-(A5):

(1) Assume that $\big(Z(t)\big)$ is a Feller process. If $\zeta^i(z, \xi)$, $i = 1, \ldots, d$, are continuous in $z$ then (A1) is satisfied. Indeed, let $\phi:(0, \infty)^d \times E \to \er$ be continuous bounded. Define
$$
g(s, z, \xi) = \int_E \phi\big(s^1\zeta^1(\tl z, \xi), \ldots,
s^d \zeta^d(\tl z, \xi), \tl z\big) P(z, d\tl z).
$$
It is continuous by the Feller property of $\big(Z(t)\big)$. Consequently, the mapping
$$
(s, z) \mapsto \ee^{(s,z)} \phi\big( S(1), Z(1) \big)
= \int_{E^\xi} g(s, z, \xi) \nu(d\xi),
$$
where $\nu$ is a distribution of $\xi(1)$ on $E^\xi$, is continuous by dominated convergence theorem and (A1) holds. In particular, if $\big(Z(t)\big)$ is a Markov chain with a finite state space (A1) is always satisfied.

(2) Assumption (A2) reads that the expected one period growth rate is finite.

(3) Assume that $\zeta^i(z, \xi)$, $i = 1, \ldots, d$, are bounded functions separated from 0 and continuous in $z$. Clearly, $h(\pi, z)$ is bounded. By (A1) $\big(Z(t)\big)$ is a Feller process, hence $h(\pi, z)$ is continuous and (A2) holds.

(4) By Jensen's inequality
$$
\inf_{\pi \in \es} h(z, \pi) = \min_{i=1, \ldots, d}
\ee^z\big\{ \ln \zeta^i \big( Z(1), \xi(1) \big) \big\}.
$$
Therefore, $h(\pi, z)$ is bounded from below if and only if
$$
\inf_{z \in E} \ee^z\big\{ \ln \zeta^i \big( Z(1), \xi(1) \big) \big\} >
-\infty, \frs i=1, \ldots, d.
$$

(5) Condition (A2) does not imply boundedness of $\zeta^i$.
Consider a generalized Black-Scholes model with economic factors (see \cite{Bielecki2}, \cite{Bielecki1}, \cite{Palczewski1}), i.e.
\begin{multline*}
S^i(t+1) = S^i(t) \exp \Big( \sigma^i\big(Z(t+1)\big)\cdot \big(W(t+1) - W(t)\big)
+ \mu^i\big( Z(t+1) \big)\Big),\\
i=1, \ldots, d,
\end{multline*}
where $\big(Z(t)\big)$ is a Feller process, $\big(W(t)\big)$ is an $m$-dimensional Wiener process and $\sigma^i: E \to  \er^m$, $\mu^i: E \to \er$, $i=1, \ldots, d$, are continuous bounded functions.
Clearly, (A1) is satisfied by (1). To show (A2) we recall the definition
$$
h(\pi, z) = \ee^z \ln \left(\sum_{i=1}^d \pi^i
\exp \Big( \sigma^i\big(Z(1)\big)\cdot\xi(1)
+ \mu^i\big( Z(1) \big)\Big)\right)
$$
with $\xi(1) = W(1) - W(0)$.
Consequently,
\begin{align*}
&\ee^z \big\{ - D_1\big(Z(1)\big) \|\xi(1)\|_2
- D_2\big(Z(1)\big) \big\}\\
&\mop{120}\le
h(\pi, z)
\le
\ee^z \big\{ D_1\big(Z(1)\big) \|\xi(1)\|_2
+ D_2\big(Z(1)\big) \big\},
\end{align*}
where $\xi$ has a standard normal distribution on $\er^m$, denoted by $\nu$,
$D_1(z) = \linebreak \max_{i=1, \ldots, d} \|\sigma^i(z)\|_2$,
$D_2(z) = \max_{i=1, \ldots, d} |\mu^i(z)|$,
and $\|\cdot\|_2$ stands for the $L^2$ norm in $\er^m$. Therefore, $h(\pi, z)$ is bounded. Continuity follows by dominated convergence theorem.

(6) In the stochastic control literature a one-step uniform ergodicity is usually assumed, i.e. (A3) with $n=1$ (see e.g. condition (UE) in \cite{Stettner2}). Allowing for $n > 1$ opens a new class of applications. In particular, 
if $Z(t)$ is a recursive Markov chain on a finite state space then (A3) holds with some $n > 0$, but usually it does not hold with $n=1$.

(7) Assumption ($\text{H}^*$): there is a continuous function $u_0$ defined on
$E$ such that $u_0(x) \geq 1$ for $x\in E$, $P u_0(x)$ is bounded on
compact subsets of $E$ and for any $l$ the set $\left\{z:
\frac{u_0(z)}{Pu_0(z)}\leq l\right\}$ is compact.

\bl
If  $E^\xi$ is locally compact and ($\text{H}^*$) is satisfied then
(A4) holds.
\el
\dowod
Without loss of generality we may assume that the support of $\nu$ is not compact (otherwise we can replace $E^\xi$ by a compact set). Let $(K_n)$ be an increasing sequence of
compact sets such that $\nu(K_{n+1}\setminus K_n)\leq \frac{1}{n^2}$, and $\overline{K_{n+1}\setminus K_n}\cap K_{n-1}=\emptyset$, for $n=1,2,\ldots$, and $\bigcup_n K_n=E^\xi$.
Define a function $g$ on $E^\xi$ to be equal to $1$ on $K_1$ and
$\sqrt{n}$ on $\overline{K_{n+1}\setminus K_n}$ for odd $n$, and
extend $g$ using Tietze theorem to a continuous function on the whole $E^\xi$.
The construction in Tieze theorem implies that $g(\xi)\geq 1$ and $\nu(g) := \int_{E^\xi}g(\xi)\nu(d\xi)<\infty$. Let $\hat{u}_0(z,\xi)=u_0(z)g(\xi)$. We shall prove that the set
$$
\Gamma_l= \bigg\{x \in \hat E: \frac{\hat u_0(x)}{\hat P \hat u_0(x)}\leq l \bigg\}
= \left\{(z,\xi) \in \hat E: \frac{u_0(z)g(\xi)}{Pu_0(z)\, \nu(g)}\leq l \right\}
$$
is compact for any $l$. Let $(z_n,\xi_n) \subset \Gamma_l$. If $(\xi_n)$ leaves all compact sets $K_m$ then $g(\xi_n) \to \infty$. Consequently $\frac{u_0(z_n)}{Pu_0(z_n)} \to 0$, which contradicts $\inf_{z\in E}\frac{u_0(z)}{Pu_0(z)}>0$. Therefore, there exists $m$ such that $(\xi_n)$ is contained in $K_m$. Compactness of $K_m$ implies that $\xi_{n_k}\to \xi\in K_m$ for some subsequence $n_k$.  Since
$$
\frac{u_0(z_{n_k})}{Pu_0(z_{n_k})}
\leq
\frac{l \nu(g)}{g(\xi_{n_k})}
\leq
l \nu(g),
$$
then by ($\text{H}^*$) there is a subsequence of $z_{n_k}$ convergent to $z$. Due to continuity
of $\frac{\hat u_0}{\hat P \hat u_0}$, we have $(z,\xi)\in \Gamma_l$, which completes the proof of compactness of $\Gamma_l$.
\qed

\section{Discounted functionals and estimates}
\label{sec:c}
This section is devoted to an in-depth study of the discounted functional related to the functional (\ref{eqn:a06}). It plays a major role in the derivation of the Bellman inequality for our optimization problem.

Given $\pi_-, x_-, z$ consider a discounted functional
\begin{multline}
J^{\pi_-, x_-, z}_\beta(\Pi) = \ee^z \bigg\{ \sum_{t=0}^\infty \beta^t h\big(\pi(t), Z(t)\big)
+ \sum_{k=1}^\infty \beta^{\tau_k} \ln e\big(\pi_-(\tau_k), \pi_k, X_-(\tau_k)\big) \bigg\},\\
\beta \in (0,1),
\label{eqn:c03}
\end{multline}
and its value function
$$
v_\beta(\pi_-, x_-, z) = \sup_{\Pi \in \ga^z} J^{\pi_-, x_-, z}_\beta(\Pi).
$$

Denote by $M$ an impulse operator acting on measurable functions
\begin{equation}
\label{eqn:c02}
M w(\pi_-, x_-, z) = \sup_{\pi \in \es} \Big\{ \ln e(\pi_-, \pi, x_-)
+ w \big(\pi, x_- \ e(\pi_-, \pi, x_-), z\big) \Big\}.
\end{equation}

\bl
\label{lem:c01}
The impulse operator maps the space of continuous bounded functions into itself. Moreover, given any bounded continuous function $w$ there exists a measurable selector for $Mw$.
\el
\dowod
The proof is standard (see  \cite{brown} Corollary 1 or \cite{Hernandez}).
\qed

\bt
\label{thm:c01}
Under (A1)-(A2) the function $v_\beta$ is continuous and bounded, and satisfies Bellman equation
\begin{align}
\label{eqn:c01}
&v_\beta(\pi_-, x_-, z) \\
\notag
&\mop{30}= \sup_{\tau}\ \ee^z \Big\{ \sum_{t=0}^{\tau - 1} \beta^t h(\pi(t), Z(t))
+ \beta^{\tau} M v_\beta\Big( \pi_-(\tau),  X_-(\tau), Z(\tau) \Big) \Big \},
\end{align}
where
\begin{align*}
\pi_-(0) = \pi_-, \qquad &\pi_-(t+1) = \pi_-(t) \diamond \zeta(t+1),\\
X_-(0) = x_-,\qquad &X_-(t+1) = X_-(t)\ \big(\pi_-(t) \cdot \zeta(t+1)\big)
\end{align*}
are counterparts of (\ref{eqn:a09}), (\ref{eqn:a11}), (\ref{eqn:a04}).
\et
\dowod
By Lemma \ref{lem:a01} the function $\ln e(\pi_-, \pi, x_-)$ is bounded, by (A2) $h(\pi, z)$ is bounded. Therefore, $v_\beta(\pi_-, x_-, z)$ is bounded. For a continuous bounded function  $v: \es \times (0, \infty) \times E \mapsto \er$ let
$$
\te_\beta v(\pi, x, z) =
\sup_{\tau}\ \ee^z \Big\{ \sum_{t=0}^{\tau - 1} \beta^t h(\pi(t), Z(t))
+ \beta^{\tau} M v\Big( \pi_-(\tau),  X_-(\tau), Z(\tau) \Big) \Big \}.
$$
The operator $\te_\beta$ maps the space $C^b = C^b(\es \times (0, \infty) \times E; \er)$ of bounded continuous functions into itself. It results from the Feller property (A1) of the transition operator of the process $\big(S(t), Z(t)\big)$ by a general result on the continuity of the value function of optimal stopping problems. Let
$$
v^0_\beta (\pi_-, x_-, z) = \sum_{t=0}^\infty \beta^t \ee^z h\big(\pi_-(t), X_-(t)\big).
$$
Put
$v^{k+1}_\beta = \te_\beta v^k_\beta$. Thanks to continuity of $v^k_\beta$ and $Mv^k_\beta$ it can be shown that $v^k_\beta$ is a value function
for the maximization of $J_\beta$ over portfolios with at most $k$ transactions. Observe that it is never optimal to have two transactions at the same time ($\prob(\tau_k = \tau_{k+1}) > 0$) by the subadditivity of the transaction costs structure. Therefore, we have the estimate
$$
\norma{v_\beta - v^k_\beta}_\infty \le \sum_{l=k}^\infty \beta^l \norma{h}_\infty
= \beta^k \frac{\norma{h}_\infty}{1 - \beta},
$$
which implies that $v^k_\beta$ tends uniformly to $v_\beta$. Consequently, $v_\beta$ is a continuous bounded function and satisfies $v_\beta = \te_\beta v_\beta$, which is equivalent to the Bellman equation (\ref{eqn:c01}).
\qed

There are two distinct cases: $C > 0$ (fixed plus proportional transaction costs) and
$C=0$ (proportional costs only). Theorem \ref{thm:c01} applies to both of them. The rest of this section is devoted to estimation of the difference between value functions of problems with and without fixed term in  transaction costs. In the beginning let us examine the equation for $e(\pi_-, \pi, x_-)$:
$$
c\big(\pi_-, e(\pi_-, \pi, x_-) \pi\big) + \frac{C}{x_-} + e(\pi_-, \pi, x_-) = 1.
$$
Clearly, if $C=0$, the solution is independent of $x_-$. Similarly, while $C=0$, the impulse operator $M$ does not depend on $x_-$, hence the value function $v_\beta$ is independent of $x_-$. We shall therefore refer to the case without a fixed term in transaction costs by skipping $x_-$ in the list of arguments and writing $\tl J^{\pi_-, z}_\beta(\Pi)$, $\tl v_\beta(\pi_-, z)$ and $\tl e(\pi_-, \pi)$.

\subsection{Technical estimates}
\label{subsec:b01}\label{subsec:b02}
This subsection presents auxiliary results. They are similar to those obtained in \cite{Palczewski2}.  For completeness, their proofs are included in Appendix.

Due to self-financing of portfolios, transaction costs decrease portfolio wealth. It is therefore important to derive estimates on the diminution factor $e(\pi_-, \pi, x_-)$ and to study the relationship between $e(\pi_-, \pi, x_-)$ and $\tl e(\pi_-, \pi)$. First of the following lemmas states lower bounds for $e$ and $\tl e$:
\bl
\label{lem:b01}
We have
\begin{align*}
1 - \tl e(\pi_-, \pi) &\le \frac{2 \max_{i}(c^1_i, c^2_i)}{1 - \max_i(c^1_i, c^2_i)},\\[2pt]
1 - e(\pi_-, \pi, x_-) &\le \frac{2 \max_{i}(c^1_i, c^2_i) + \frac{C}{x_-}}{1 - \max_i(c^1_i, c^2_i)}.
\end{align*}
\el

Let $x^* = \inf\{ x_- :\ e(\pi_-, \pi, x_-) > 0 \text{ for all } \pi_-, \pi \in \es\}$. If the wealth of the portfolio is greater than $x^*$, any transaction can be executed. We use this rough treshold in the following lemma:
\bl
\label{lem:d01}
For $\pi_-, \pi \in \es$
\begin{itemize}
\item[i)] $\displaystyle
e(\pi_-, \pi, \tl x_-) \le e(\pi_-, \pi, x_-) \le \tl e(\pi_-, \pi), \frs x_- \ge \tl x_- > 0$.
\item[ii)] $\displaystyle
\tl e(\pi_-, \pi) - e(\pi_-, \pi, x_-) \le \frac{C}{  \big(1 - \max_{i} c^1_i \big) x_-} \quad$ for $x_- > x^*$.
\item[iii)] For all $M > x^*$ and $x_- \ge M$
$$
\displaystyle
\ln \frac{\tl e(\pi_-, \pi)}{e(\pi_-, \pi, x_-)} \le
\frac{1}{\inf_{\tl \pi_-, \tl \pi} e(\tl \pi_-, \tl \pi, M)}
\ \frac{C}{
\big(1 - \max_{i} c^1_i\big) x_-}.
$$ 
\end{itemize}
\el

\bw
\label{col:d01}
The value function $v_\beta(\pi_-, x_-, z)$ is non-decreasing in $x_-$.
\ew

Due to Theorem \ref{thm:c01} the value function $\tl v_\beta(\pi_-, z)$ is bounded and continuous for each $\beta$.
However, it does not imply that it is uniformly bounded in $\beta$. Conversely, it increases to infinity as $\beta$ grows to $1$ in market models of interest. To account for this fact, we shall study the span seminorm of $\tl v_\beta$, which is defined as
$\norma{\tl v_\beta}_{sp} = \sup \tl v_\beta (\cdot) - \inf \tl v_\beta (\cdot)$. 
\bl
\label{lem:c02}
Under (A3) there exists $M < \infty$ such that
$$
\norma{\tl v_\beta}_{sp} \le M,
$$
for all $\beta \in (0,1)$.
\el

\subsection{Large deviations and proportional transaction costs}
Theorem \ref{lem:b02} provides important insight into the dynamics of asset prices. In this section we apply this result to describe the dynamics of the portfolio wealth under proportional transaction costs. Let us introduce a general assumption:
\begin{opisanie}{(A6)}
$\eta < \hat p$,
\end{opisanie}
 where $\hat p$ is a constant from Theorem \ref{lem:b02} and
$$
\eta = - \ln \bigg(1 - \frac{2 \max_i(c^1_i, c^2_i)}{1-\max_i(c^1_i, c^2_i)}\bigg).
$$
Since $\eta$ is a unique solution to the equation
$$
e^{-\eta} = 1 - \frac{2 \max_i(c^1_i, c^2_i)}{1-\max_i(c^1_i, c^2_i)},
$$
by virtue of Lemma \ref{lem:b01}, $e^{-\eta}$ is a lower bound on $\tl e(\pi_-, \pi)$.
Formula (\ref{eqn:a04}) gives the following estimate
$$
X_-(t) \ge X_-(0) \prod_{s=0}^{t-1} \Big( e^{-\eta}\ \pi(s) \cdot \zeta(s+1)\Big)
= X_-(0)  e^{-\eta t} \prod_{s=0}^{t-1} \Big(\pi(s) \cdot \zeta(s+1)\Big).
$$
Consequently, denoting $\hat \zeta(t) = \min\big(\zeta^1(t), \ldots, \zeta^d(t)\big)$, we obtain
\begin{equation}
\label{eqn:b02}
X_-(t) \ge X_-(0)  e^{-\eta t} \prod_{s=0}^{t-1} \hat \zeta(s+1).
\end{equation}
In view of Theorem \ref{lem:b02} for any $\epsilon > 0$ there exists $K>0$, $\gamma > 0$ and $T^*$ such that
$$
\prob \Big\{ e^{T(\hat p - \epsilon)} \le \prod_{s=0}^{t-1} \hat \zeta(s+1) \Big\} \ge
1 - K e^{-\gamma T}, \quad T > T^*,
$$
and the wealth of the portfolio satisfies
$$
\prob \Big\{  X_-(0) e^{T(\hat p -\eta - \epsilon)} \le X_-(T)\Big\} \ge
1 - K e^{-\gamma T}, \quad T > T^*.
$$
Due to (A6) there exists $0 < \epsilon < \hat p - \eta$, which implies that $X_-(t)$ increases exponentially fast irrespective of portfolio trading strategy.

\subsection{Bounds on $\tl v_\beta(\pi_-, z) - v_\beta(\pi_-, x_-, z)$}
\label{subsec:c3}
For the rest of this section assume that (A1)-(A6) are satisfied.
Since
$$
\lim_{m \to \infty} - \ln \bigg(1 - \frac{2 \max_i(c^1_i, c^2_i) + \frac{C}{m}}{1-\max_i(c^1_i, c^2_i)}\bigg) = \eta < \hat p,
$$
there exists a constant $M > 0$ such that
\begin{equation}
\label{eqn:b03}
\hat p > \eta_M := - \ln \bigg(1 - \frac{2 \max_i(c^1_i, c^2_i) + \frac{C}{M}}{1-\max_i(c^1_i, c^2_i)}\bigg).
\end{equation}
\bt
\label{thm:b02}
For any $z \in E$ and any admissible strategy $\tl \Pi \in \laa^z$, $\pi_- \in \es$, $x_- \in (0, \infty)$ there exists an admissible trading strategy $\Pi \in \laa^z$ such that
$$
 \tl J_\beta^{\pi_-, z}(\tl \Pi) - J_\beta^{\pi_-, x_-, z}(\Pi) \le L(x_-), \quad \beta \in (0,1),
$$
where
$$
L(x_-) = K_1 + K_2 \max(K_3, -\ln x_-)
$$
for some strictly positive constants $K_1, K_2, K_3$ independent of the choice of $z$, $\tl \Pi$, $\pi_-$ and $x_-$.
\et
\bw
\label{cor:b01}
We have
$$
0 \le \tl v_\beta(\pi_-, z) - v_\beta(\pi_-, x_-, z) \le L(x_-), \quad \beta \in (0,1),\  \pi_- \in \es,\ z \in E.
$$
where $L(x_-)$ is a function from Theorem \ref{thm:b02}.
\ew
\dowod
The inequality $0 \le \tl v_\beta(\pi_-, z) - v_\beta(\pi_-, x_-, z)$ is obvious. For the second inequality it is enough to notice that
$$
\tl v_\beta(\pi_-, z) - v_\beta(\pi_-, x_-, z) \le
\sup_{\tl \Pi \in \ga}\Big\{ \tl J_\beta^{\pi_-, z}(\tl \Pi) - \tl J_\beta^{\pi_-, x_-,  z}(\Pi) \Big\},
$$
where by $\Pi$ we denote a strategy related to $\tl \Pi$ as in Theorem \ref{thm:b02}.
\qed

The strength of the above theorem and corollary lies in the fact that the estimates are uniform in $\beta$, $\pi_-$ and $z$.

\dowod[of Theorem \ref{thm:b02}]
Fix $\pi_-, x_-, z$ and $\tl \Pi \in \laa^z$. We construct $\Pi$, with its pre-transaction wealth denoted by $X_-(t)$, in the following way: if $X_-(t) \ge M$ we mimic the strategy $\tl \Pi$, i.e. we keep the same proportions of stocks. On the other hand, if $X_-(t)$ is smaller than $M$ we do not make any transactions and wait for the wealth to raise over $M^* = M e^{\eta_M}$. At that moment we perform a transaction to make the proportions equal to those defined by $\tl \Pi$. This decreases the wealth at most by $e^{-\eta_M}$, so the resulting portfolio wealth is not less than $M$.

Let $\tl \pi_-(t)$ and $\tl \pi(t)$ denote the pre-transaction and the post-transaction process linked to the strategy $\tl \Pi$. Analogously, $\pi_-(t)$ and  $\pi(t)$ are the processes corresponding to the strategy $\Pi$.
By the construction of $\Pi$ we know that $\pi(t) = \tl \pi(t)$ if $X_-(t) \ge M^*$. However, if the wealth $X_-(t)$ is below $M^*$ but above $M$  we cannot determine whether $\pi(t) = \tl \pi(t)$. This is caused by the fact that the wealth $X_-(t)$ can be between $M$ and $M^*$ as a result of either normal investing process or recovering from the shortage of wealth (being below $M$). 

By the definition of $J_\beta^{\pi_-, x_-, z}$ and $\tl J_\beta^{\pi_-, z}$ we have
\begin{align*}
\tl J_\beta^{\pi_-, z}&(\tl \Pi) - J_\beta^{\pi_-, x_-, z}(\Pi)\\
&=  \ee^z \bigg\{ \sum_{t=0}^\infty \beta^t
\bigg(
 h\big(\tl \pi(t), Z(t)\big) - h\big(\pi(t), Z(t)\big)\\
& \mop{75} + \ind{\tl \pi_-(t) \ne \tl \pi(t)}  \ln \tl e\big(\tl \pi_-(t), \tl \pi(t)\big)\\
& \mop{75} - \ind{\pi_-(t) \ne \pi(t)}  \ln e\big(\pi_-(t), \pi(t), X_-(t)\big)
\bigg) \bigg\}.
\end{align*}
Above difference can be bounded from above by the sum of the following two expressions:
\begin{align}
\label{eqn:b05}
 &\ee^z \bigg\{ \sum_{t=0}^\infty \beta^t
\ind{X_-(t) < M^*} \bigg(
 h\big(\tl \pi(t), Z(t)\big) - h\big(\pi(t), Z(t)\big)\\
&\notag\mop{160}-   \ind{\pi_-(t) \ne \pi(t)} \ln e\big(\pi_-(t), \pi(t), X_-(t)\big)
\bigg) \bigg\},\\
\label{eqn:b06}
&\ee^z \bigg\{ \sum_{t=0}^\infty \beta^t
\ind{X_-(t) \ge M^*} \bigg(
 h\big(\tl \pi(t), Z(t)\big) - h\big(\pi(t), Z(t)\big)\\
&\notag\mop{160}+   \ind{\pi_-(t) \ne \pi(t)} \ln \frac{\tl e\big(\pi_-(t), \pi(t)\big)}{e\big(\pi_-(t), \pi(t), X_-(t)\big)}
\bigg) \bigg\}.
\end{align}
By construction of the strategy $\Pi$ no transaction is performed if the wealth $X_-(t)$ is below $M$, so we
have
$$
- \ind{\pi_-(t) \ne \pi(t)} \ln e\big(\pi_-(t), \pi(t), X_-(t)\big) \le \eta_M.
$$
This yields
$$
(\ref{eqn:b05}) \le L_1\  \ee^z \sum_{t=0}^\infty \ind{X_-(t) < M^*},
$$
where $L_1 = \sup h(\cdot) - \inf h(\cdot) + \eta_M$. On the other hand, if $X_-(t) \ge M^*$, we have
$\pi(t) = \tl \pi(t)$, so
$$
(\ref{eqn:b06}) \le \ee^z \bigg\{ \sum_{t=0}^\infty \bigg(
\ind{X_-(t) \ge M^*} \ln \frac{\tl e\big(\pi_-(t), \pi(t)\big)}{e\big(\pi_-(t), \pi(t), X_-(t)\big)}
\bigg)\bigg\}.
$$
By virtue of Lemma \ref{lem:d01} (iii) we obtain
$$
(\ref{eqn:b06}) \le \ee^z \bigg\{ \sum_{t=0}^\infty \bigg(
\ind{X_-(t) \ge M^*} \frac{L_2}{X_-(t)}\bigg)\bigg\},
$$
where
$$
L_2 = \frac{C}{\inf_{\hat \pi_-, \hat \pi} e(\hat \pi_-,\hat \pi, M^*)}.
$$
Consequently, we have the estimate
\begin{multline}
\label{eqn:b07}
\tl J_\beta^{\pi_-, z}(\tl \Pi) - J_\beta^{\pi_-, x_-, z}(\Pi) \\
\le
L_1\  \ee^z \sum_{t=0}^\infty \ind{X_-(t) < M^*} +
L_2 \ \ee^z \sum_{t=0}^\infty \bigg(
\ind{X_-(t) \ge M^*} \frac{1}{X_-(t)}\bigg).
\end{multline}

To complete the proof we use the large deviations estimate. Fix $\epsilon > 0$ small enough so that $\hat p - \eta_M - \epsilon > 0$. Denote by $A_t$ the event
$$
A_t = \Big\{ \frac{1}{t} \ln\Big(\prod_{j=0}^{t-1} \hat \zeta\big(Z(j+1), \xi(j+1)\big)\Big) - \hat p \ge -\epsilon \Big\}.
$$
The strategy $\Pi$ is constructed in such a way that trade takes place only if $X_-(t) \ge M$. Thus on the set $A_t$ we have
\begin{equation}
\label{eqn:b08}
X_-(t) \ge x_- e^{-t \eta_M} e^{t(\hat p - \epsilon)} = x_- e^{t(\hat p - \eta_M - \epsilon)}.
\end{equation}
This reads as an exponentially fast growth of the wealth due to $\hat p - \eta_M - \epsilon > 0$.

Let $K>0$, $\gamma > 0$, and $T^*$ be the constants from Theorem \ref{lem:b02} (iii) for the given $\epsilon$. By the large deviations estimate we have
$$
\prob^z (A^c_t) \le K e^{-\gamma t}\quad \text{for } t \ge T^*,
$$
where by $A^c_t$ denotes the complement of $A_t$. Let $t_0$ be the smallest integer such that $t_0 \ge T^*$ and
$$
e^{t_0(\hat p - \eta_M - \epsilon)} \ge \frac{M^*}{x_-}.
$$
Clearly, $X_-(t) \ge M^*$ on $A_{t}$ for all $t \ge t_0$. Hence
$$
\ee^z \sum_{t=0}^\infty \ind{X_-(t) < M^*} \le
t_0 + \sum_{t=t_0}^\infty \prob(A^c_t) \le t_0 + \sum_{t=t_0}^\infty K e^{-\gamma t} =: L_3.
$$
Computation of a bound for the second term of (\ref{eqn:b07}) has to be split into two parts depending on $A_t$:
\begin{align*}
\ee^z \sum_{t=0}^\infty \bigg(&
\ind{X_-(t) \ge M^*} \frac{1}{X_-(t)}\bigg)\\
&= \ee^z \sum_{t=0}^\infty \bigg(
\ind{X_-(t) \ge M^*} \frac{ \ind{A_t}}{X_-(t)} \bigg)
+
\ee^z \sum_{t=0}^\infty \bigg(
\ind{X_-(t) \ge M^*} \frac{\ind{A_t^c}}{X_-(t)} \bigg).
\end{align*}
Easily,
\begin{align*}
\ee^z \sum_{t=0}^\infty \bigg(& 
\ind{X_-(t) \ge M^*} \frac{\ind{A_t^c}}{X_-(t)} \bigg)\\
&\le
\frac{1}{M^*} \ee^z \sum_{t=0}^\infty \prob(A^c_t) \le
\frac{1}{M^*} \ee^z \sum_{t=0}^\infty K e^{-\gamma t} =: L_4.
\end{align*}
Due to (\ref{eqn:b08})
\begin{align*}
\ee^z \sum_{t=0}^\infty \bigg(&
\ind{X_-(t) \ge  M^*} \frac{\ind{A_t}}{X_-(t)} \bigg)\\
&\le
\ee^z \sum_{t=0}^{t_0-1} \bigg(
\ind{X_-(t) \ge M^*} \frac{\ind{A_t}}{X_-(t)} \bigg)
+
\ee^z \sum_{t=t_0}^\infty \bigg(
\ind{X_-(t) \ge M^*} \frac{\ind{A_t}}{X_-(t)} \bigg)\\
&\le
\frac{t_0}{M^*} +
\frac{1}{M^*} \sum_{t=t_0}^\infty e^{-(t-t_0)(\hat p - \eta_M - \epsilon)} \\
&=: L_5.
\end{align*}
Consequently,
$$
\tl J_\beta^{\pi_-, z}(\tl \Pi) - J_\beta^{\pi_-, x_-, z}(\Pi) \le
L_1 L_3 + L_2 (L_4 + L_5).
$$
The constants $L_1, \ldots, L_5$ do not depend on $\pi_-$, $\tl \Pi$ and $z$. However, they depend on $x_-$ through $t_0$. Combining the estimates for $L_1, \ldots, L_5$ we obtain the formula for $L(x_-)$.
\qed

\section{Growth optimal portfolios}
\label{sec:d}
Now we are in a position to state and prove the main result of this paper: existence and form of an optimal strategy maximizing the expected average rate of return of a portfolio of financial assets.
\bt
\label{thm:b01}
Under assumptions (A1)-(A6) there exists a measurable function $p: \es \times (0, \infty) \times E \to \es$, a constant $\lambda$ and a measurable set $I \subseteq \es \times (0, \infty) \times E$ such that
\begin{equation}
\label{eqn:b01}
\lambda = J^{\pi_-, x_-, z} (\Pi^*) = \sup_{\Pi \in \ga^z} J^{\pi_-, x_-, z} (\Pi),
\end{equation}
where the optimal portfolio $\Pi^* = \big((\pi^*_1, \tau^*_1), (\pi^*_2, \tau^*_2), \ldots\big)$ is given by the formulas
\begin{align*}
&\tau^*_1 = \inf \{ t \ge 0 \setcond \big(\pi_-(t), X_-(t), Z(t)\big) \in I \},\\
&\tau^*_{k+1} = \inf \{ t > \tau^*_k \setcond \big(\pi_-(t), X_-(t), Z(t)\big) \in I \},\\
&\pi^*_k = p\big(\pi_-(\tau^*_k), X_-(\tau^*_k), Z(\tau^*_k)\big).
\end{align*}
\et
The strength of the above theorem is in its generality. We are not aware of papers dealing with the maximisation of the average rate of return in such a general setting and with fixed and proportional transaction costs.  This result also extends the area of applicability of the vanishing discount approach to models with non-weakly continuous controlled transition probabilities. Existing results require either strongly or weakly continuous (Feller) controlled transition probabilities (see \cite{Hernandez}, \cite{jaskiewicz}, \cite{Schal}, \cite{Stettner}). Moreover, in Section \ref{sec:g} we generalize Theorem \ref{thm:b01} to other transaction costs structures.
\begin{wn}
\label{cor:c01}{\ }
\begin{itemize}
\item [i)]
The optimal value for the problem with only proportional transaction costs ($C=0$) is equal to $\lambda$ from Theorem \ref{thm:b01}. The strategy optimal for fixed plus proportional transaction costs is also optimal for proportional transaction costs.
\item [ii)]
There exists an optimal portfolio $\Pi$ for the problem with proportional transaction costs that depends only on the current state of the processes $(\pi_-(t))$ and $(Z(t))$ (does not depend on $(X_-(t))$). 
\item [iii)]
If $\Pi$ is the portfolio from (ii), then the portfolio $\Pi_M$ optimal for 
fixed plus proportional transaction costs is constructed as follows (for notation consult Subsection \ref{subsec:c3}): whenever $X_-(t)<M$ do not make any transactions and wait until the wealth increases over $Me^{\eta_M}$; otherwise as long as $X_-(t) \geq M$ keep the same proportions of stocks as in $\Pi$. 
\end{itemize}
\end{wn}
Optimal strategies for proportional costs can be efficiently computed in a number of cases: there are closed-form formulas in simple  diffusion models and efficient algorithms for more complicated models, all benefiting from compactness of the state space. Corollary \ref{cor:c01} (iii) presents how an optimal portfolio for proportional cost can be employed in construction an optimal trading strategy for fixed and proportional transaction costs. However, unlike portfolio $\Pi^*$ from Theorem \ref{thm:b01}, portfolio $\Pi_M$ constructed in Corollary \ref{cor:c01} depends on past variations of its wealth and hence is not Markovian. 
The proof of Corollary \ref{cor:c01} is presented in details later.
\bigskip

\dowod[of Theorem \ref{thm:b01}]
We use a generalization of the vanishing discount method (\cite{Arapost}, \cite{Hernandez}, \cite{jaskiewicz}, \cite{Schal}, \cite{Stettner2}). We obtain a Bellman inequality for our optimization problem as a limit of Bellman equations for discounted problems (\ref{eqn:c03}). We cannot directly apply known results since they require continuity of the controlled transition function $q$ defined below. Instead, we follow the approach pioneered by \cite{Schal} exchange the parts where the continuity of $q$ is needed by considerations based on specific properties of our control problem. We also ease the requirement of local compactness of the state space in the spirit of \cite{jaskiewicz}.

Denote by $\eh = \es \times (0,\infty) \times E$ the state space of our Markovian control model. It is complete and separable, which is needed for the existence of measurable selectors. Denote by $q$ the controlled transition operator, i.e. a function $q : \eh \times \es \to \pe(\eh)$, where $\pe(\eh)$ is the space of Borel probability measures on $\eh$, uniquely determined by the formula
\begin{multline}
\label{eqn:e01}
\int_\ha f(\tl \pi_-, \tl x_-, \tl z)\ q(\pi_-, x_-, z, \pi)(d\tl\pi_-, d\tl x_-, d\tl z)\\
= \ee^z f\big(\pi \diamond \zeta\big(z, \xi(1)\big), X^\pi_-(1), z(1) \big)
\end{multline}
for all bounded measurable $f : \eh \to \er$,
where
$$
X^\pi_-(1) = \begin{cases}
x_- \; e(\pi_-, \pi, x_-)\ \big(\pi \cdot \zeta\big(z, \xi(1)\big)\big),
& \text{ when }\pi_- \ne \pi,\\
x_-\; \big(\pi \cdot \zeta\big(z, \xi(1)\big)\big),& \text{ when } \pi_- = \pi.
         \end{cases}
$$
Obviously, $q$ is not weakly continuous as long as the constant term
in transaction costs is non-null. Indeed, $X^{\pi_-}_-(1) - X^{\tl \pi}_-(1) \ge C$
for any $\tl \pi \ne \pi_-$. Consider
$$
\eta (\pi_-, \pi, x_-, z) = \begin{cases}
h(\pi, z), & \pi_- = \pi,\\
h(\pi, z) + \ln e(\pi_-, \pi, x_-), & \pi_- \ne \pi.
                         \end{cases}
$$
Bellman equation (\ref{eqn:c01}) writes in an equivalent form
\begin{equation}
\label{eqn:e02}
v_\beta(\pi_-, x_-, z) = \sup_{\pi \in \es}\ \Big\{\eta(\pi_-, \pi, x_-, z)
+ \beta \int v_\beta\ dq(\pi_-, x_-, z, \pi) \Big\}.
\end{equation}
Let $a_\beta:\eh \to \es$ be a measurable selector
for $Mv_\beta$ (see Lemma \ref{lem:c01}) and $I_\beta$ be the impulse region
$$
I_\beta = \{ (\pi_-, x_-, z) \in \eh \setcond v_\beta(\pi_-, x_-, z) = M v_\beta(\pi_-, x_-, z) \}.
$$
The optimal strategy in this formulation is given by a measurable function
$f_\beta: \eh \to \es$
$$
f_\beta(\pi_-, x_-, z) =
\begin{cases}
\pi_-, & (\pi_-, x_-, z) \notin I_\beta,\\
a_\beta(\pi_-, x_-, z), & (\pi_-, x_-, z) \in I_\beta.
\end{cases}
$$

Since $v_\beta$ is unbounded as $\beta$ grows to $\infty$ we introduce the relative discounted value function
$$
w_\beta(\pi_-, x_-, z) = m_\beta - v_\beta(\pi_-, x_-, z),
$$
where
$$
m_\beta = \sup_{\pi_- \in \es}\; \sup_{z \in E}\; \tl v_\beta(\pi_-, z)
$$
is well-defined due to Lemma \ref{lem:c02}.
Moreover, we have
\bl
\label{lem:e01}
$ $
\begin{itemize}
\item[i)] $\displaystyle 0 \le w_\beta(\pi_-, x_-, z) \le M_1 + M_2 \max(M_3, -\ln x_-)$ with
$M_1, M_2, M_3 > 0$ independent of $\beta, \pi_-, x_-, z$.
\item[ii)] $\{ (1-\beta) m_\beta: \beta \in (0,1) \}$ is a pre-compact set, i.e. its closure is compact.
\end{itemize}
\el
\dowod
By Lemma \ref{lem:c02}, and Corollary \ref{cor:b01} we have
$$
w_\beta(\pi_-, x_-, z)  \le
m_\beta - v_\beta(\pi_-, z) + v_\beta(\pi_-, z) - v_\beta(\pi_-, x_-, z) \le
M + L(x_-),
$$
where $L(x_-)$ is a function defined in Theorem \ref{thm:b02}. We conclude by using the form of $L(x_-)$.
Part ii) follows from boundedness of $h(\cdot)$ and $\ln \tl e(\cdot)$.
\qed

Put $\ubl = \limsup_{\beta \uparrow 1} (1-\beta) m_\beta$, which is finite by Lemma \ref{lem:e01} (ii). Denote by $\beta_k$ the sequence of discount factors converging to $1$ such that
$$
\ubl = \lim_{k \to \infty}\ (1 - \beta_k) m_{\beta_k}.
$$
Let
$$
\underline{w}(\vartheta) = \liminf_{k \to \infty,\, \vartheta' \to \vartheta} w_{\beta_k}(\vartheta'),  \exs \vartheta \in \ha.
$$
It can be written equivalently as
$$
\underline{w}(\vartheta) = \inf \big\{ \liminf_{k \to \infty} w_{\beta_k}(\vartheta_k): \vartheta_k \to \vartheta \big\},  \exs \vartheta \in \ha.
$$
\bl
\label{lem:e02}
(\cite{jaskiewicz2} Lemma 3.1)
The function $\underline{w}$ is lower semi continuous.
\el
The proof of this lemma is straightforward and is based on the following reformulation of the definition of $\underline{w}$:
$$
\underline{w}(\vartheta) = \sup_n \inf_{k \ge n} \Big\{ \inf_{\vartheta' \in B(\vartheta, 1/n)} w_{\beta_k}(\vartheta')\Big\},
$$
where $B(\vartheta, 1/n)$ is a ball in $\eh$ of radius $1/n$.

In the sequel we use two transition operators related to $q$. Let
$\underline q$ be given by the formula (\ref{eqn:e01}) with
$$
X^\pi_-(1) =
x_- \; e(\pi_-, \pi, x_-)\ \big(\pi \cdot \zeta\big(z, \xi(1)\big)\big)
$$
and $\overline q$ with
$$
X^\pi_-(1) =
x_- \; \big(\pi \cdot \zeta\big(z, \xi(1)\big)\big).
$$
They are weakly continuous. Indeed, it is straightforward by (A1) and the continuity of $e(\pi_-, \pi, x_-)$ (see Lemma \ref{lem:a01}) that the mapping
$$
(\pi_-, x_-, z) \mapsto \big(\int_\eh f\, d \underline q(\pi_-, x_-, z),
\int_\eh f\, d \overline q(\pi_-, x_-, z)\big)
$$
is continuous for any continuous bounded function $f:\eh \to \er$.

\bl
\label{lem:e03}
(\cite{Serfozo} Lemma 3.2) Let $\{\mu_n\}$ be a sequence of probability measures on a separable metric space $\mathcal X$ converging weakly to $\mu$
and $\{g_n\}$ be a sequence of measurable nonnegative functions on $\mathcal X$. Then
$$
\int \underline{g}\, d\mu \le \liminf_{n \to \infty} \int g_n\, d\mu_n,
\quad\text{ where }\quad
\underline{g}(x) = \liminf_{n \to \infty,\ y \to x} g_n(y), \frs x \in \mathcal X.
$$
\el

\bt
\label{thm:e01}
Under assumptions (A1)-(A5) there exists a measurable function $f_1: \ha \to \es$ and a measurable
function $w: \ha \to (-\infty, 0]$ such that
\begin{equation}
\label{eqn:e05}
w(\vartheta) + \ubl \le \eta\big(\vartheta, f_1(\vartheta)\big)
+ \int w(\vartheta') q\big(\vartheta, f_1(\vartheta)\big)(d \vartheta'),
\frs \vartheta \in \ha.
\end{equation}
\et
\dowod
From equation (\ref{eqn:e02}) we derive
\begin{multline*}
w_\beta(\vartheta) + (\beta - 1) m_\beta =
- \eta\big(\vartheta, f_\beta(\vartheta)\big)
+ \beta \int w_\beta(\vartheta') q\big(\vartheta, f_\beta(\vartheta)\big)(d \vartheta'),\\
\frs \vartheta \in \ha,\exs \beta \in (0,1),
\end{multline*}
where $f_\beta$ defines an optimal strategy for $v_\beta$. Fix $\vartheta \in \eh$ and a sequence $(\vartheta_k)$ converging to $\vartheta$. Above equation can be rewritten as
$$
w_{\beta_k}(\vartheta_k) + (\beta_k - 1) m_{\beta_k} =
- \eta\big(\vartheta_k, f_{\beta_k}(\vartheta_k)\big)
+ \beta_k \int w_{\beta_k}(\vartheta') q\big(\vartheta_k, f_{\beta_k}(\vartheta_k)\big)(d \vartheta').
$$
Applying $\liminf_{k \to \infty}$ on both sides yields
\begin{align}
\label{eqn:e03}
&\liminf_{k \to \infty} w_{\beta_k}(\vartheta_k) - \ubl \\
&\notag\mop{30} =
- \limsup_{k \to \infty} \eta\big(\vartheta, f_{\beta_k}(\vartheta_k)\big)
+ \liminf_{k \to \infty} \int \beta_k w_{\beta_k}(\vartheta') q\big(\vartheta_k, f_{\beta_k}(\vartheta_k)\big)(d \vartheta').
\end{align}
Since $\es$ is compact there exists a sequence $(n_k)$ such that $f_{\beta_{n_k}}(\vartheta) \to \pi^*$ and either (a) $\vartheta_{n_k} \in I_{\beta_{n_k}}$ for every $k$, or (b) $\vartheta_{n_k} \notin I_{\beta_{n_k}}$ for every $k$.
Assume first that (a) holds. By virtue of Lemma \ref{lem:e03} we have
$$
\liminf_{k \to \infty} \int \beta_k w_{\beta_k}(\vartheta') q\big(\vartheta_k, f_{\beta_k}(\vartheta_k)\big)(d \vartheta')
\ge \int \underline{w}(\vartheta') \underline{q}\big(\vartheta, \pi^*)(d \vartheta').
$$
By Corollary \ref{col:d01} the functions $v_\beta(\pi_-, x_-, z)$ are nondecreasing in $x_-$.
This implies that $\underline{w}(\pi_-, x_-, z)$ is non-increasing in $x_-$. Hence
$\int \underline{w}(\vartheta') \underline{q}\big(\vartheta, \pi^*\big)(d \vartheta') \ge
\int \underline{w}(\vartheta') q\big(\vartheta, \pi^*\big)(d \vartheta')$ and
\begin{equation}
\label{eqn:e04}
\liminf_{k \to \infty} \int \beta_k w_{\beta_k}(\vartheta') q\big(\vartheta_k, f_{\beta_k}(\vartheta_k)\big)(d \vartheta')
\ge \int \underline{w}(\vartheta') q\big(\vartheta, \pi^*\big)(d \vartheta').
\end{equation}
In the case (b) we have $f_{\beta_{n_k}}(\vartheta_k) = \pi^k_-$, where $\vartheta_k = (\pi^k_-, x^k_-, z^k)$. Obviously, $\pi^* = \pi_-$, where $\vartheta = (\pi_-, x_-, z)$. From equalities
$q\big(\vartheta, \pi^*\big) = \overline{q}\big(\vartheta, \pi^*\big)$
 and
$q\big(\vartheta_{n_k}, f_{\beta_{n_k}}(\vartheta_k)\big) = \overline{q}\big(\vartheta_{n_k}, f_{\beta_{n_k}}(\vartheta_k)\big)$
and Lemma \ref{lem:e03} we obtain (\ref{eqn:e04}).
Since $\eta$ is upper semicontinuous we conclude that
$$
\liminf_{k \to \infty} w_{\beta_k}(\vartheta_k) - \ubl \ge - \eta\big(\vartheta, \pi^*\big)
+ \int \underline{w}(\vartheta') q\big(\vartheta, \pi^*\big)(d \vartheta').
$$
Consequently,
$$
\liminf_{k \to \infty} w_{\beta_k}(\vartheta_k) - \ubl
\ge \inf_{\pi \in \es} \big\{ - \eta\big(\vartheta, \pi\big)
+ \int \underline{w}(\vartheta') q\big(\vartheta, \pi\big)(d \vartheta')\big\}.
$$
Taking infimum over all sequences $\vartheta_n$ converging to $\vartheta$ we finally obtain
\begin{equation}
\label{eqn:e06}
\underline{w}(\vartheta) - \ubl
\ge \inf_{\pi \in \es} \big\{ - \eta\big(\vartheta, \pi\big)
+ \int \underline{w}(\vartheta') q\big(\vartheta, \pi\big)(d \vartheta')\big\}.
\end{equation}
To complete the proof we have to show that there exists a measurable selector for the infimum on the right-hand side of (\ref{eqn:e06}). Corollary \ref{col:d01} implies that $\underline{w}$ is non-increasing in $x_-$. Thus, for $(\pi_-, x_-, z) \in \ha$
$$
\int \underline{w}(\vartheta') \overline{q}\big(\pi_-, x_-, z, \pi_-\big)(d \vartheta')
\le
\int \underline{w}(\vartheta') \underline{q}\big(\pi_-, x_-, z, \pi_-\big)(d \vartheta'),
$$ 
and the infinum in (\ref{eqn:e06}) can be equivalently written as
\begin{align}
\label{eqn:e07}
\min \Big\{ &- \eta\big(\vartheta, \pi_-\big)
+ \int \underline{w}(\vartheta') \overline{q}\big(\vartheta, \pi_-\big)(d \vartheta'),\\
\notag &\inf_{\pi \in \es} \big\{ - \eta\big(\vartheta, \pi\big)
+ \int \underline{w}(\vartheta') \underline{q}\big(\vartheta, \pi\big)(d \vartheta')\big\}\Big\},
\end{align}
where $\vartheta = (\pi_-, x_-, z)$. Recall that by Lemma \ref{lem:e02} the function $\underline{w}$ is lower semicontinuous. By weak continuity of the transition probabilities $\underline{q}, \overline{q}$ the mappings
\begin{align*}
&(\pi_-, x_-, z, \pi) \mapsto \int_\eh \underline{w}(\vartheta') \, \underline{q}(\pi_-, x_-, z, \pi)(d \vartheta')\\
&(\pi_-, x_-, z, \pi) \mapsto \int_\eh \underline{w}(\vartheta') \, \overline{q}(\pi_-, x_-, z, \pi)(d \vartheta')
\end{align*}
are lower semicontinuous (see \cite{gonzalez} Lemma 3.3 (a)). Corollary 1 in \cite{brown} implies that there exists a measurable selector $f_2: \ha \to \es$ for 
$$
\inf_{\pi \in \es} \big\{ - \eta\big(\vartheta, \pi\big)
+ \int \underline{w}(\vartheta') \underline{q}\big(\vartheta, \pi\big)(d \vartheta')\big\}.
$$
Define $f_1: \ha \to \es$ by
$$
f_1(\pi_-, x_-, z) =
\begin{cases}
\pi_-, &\text{if} 
\begin{aligned}
&- \eta\big(\vartheta, \pi_-\big)
+ \int \underline{w}(\vartheta') \overline{q}\big(\vartheta, \pi_-\big)(d \vartheta') \\[-4pt]
&\le
\inf_{\pi \in \es} \big\{ - \eta\big(\vartheta, \pi\big)
+ \int \underline{w}(\vartheta') \underline{q}\big(\vartheta, \pi\big)(d \vartheta')\big\},
\end{aligned}
\\
& \\
f_2(\pi_-, x_-, z), &\text{if}
\begin{aligned}
&- \eta\big(\vartheta, \pi_-\big)
+ \int \underline{w}(\vartheta') \overline{q}\big(\vartheta, \pi_-\big)(d \vartheta') \\[-4pt]
&>
\inf_{\pi \in \es} \big\{ - \eta\big(\vartheta, \pi\big)
+ \int \underline{w}(\vartheta') \underline{q}\big(\vartheta, \pi\big)(d \vartheta')\big\},
\end{aligned}
\end{cases}
$$
and put  $w = -\underline{w}$. This completes the proof.
\qed

Fix $(\pi_-, x_-, z) \in \ha$ and define a portfolio $\Pi = \big((\pi_1, \tau_1),
(\pi_2, \tau_2), \ldots\big)$ by formulas given in Theorem \ref{thm:b01} with
$I = \{(\pi_-, x_-, z) \in \ha: f_1(\pi_-, x_-, z) \ne \pi_-\}$ and $p = f_1$. Iterating
(\ref{eqn:e05}) $T$ times, dividing by $T$ and passing with $T$ to infinity we obtain
$$
\ubl \le J^{\pi_-, x_-, z}(\Pi) + \liminf_{T \to \infty} \ee^z \frac{w\big(\pi^\Pi_-(T), X^\Pi_-(T), Z(T)\big)}{T} \le J^{\pi_-, x_-, z}(\Pi),
$$
since $w$ is nonpositive. On the other hand, by a well-known Tauberian relation
\begin{align*}
J^{\pi_-, x_-, z}(\Pi) &\le \liminf_{\beta \to 1} (1-\beta) J_\beta^{\pi_-, x_-, z}(\Pi)\\
&\le \liminf_{\beta \to 1} (1-\beta) v_\beta(\pi_-, x_-, z)
\le \liminf_{\beta \to 1} (1-\beta) v_\beta(\pi_-, z) \le \ubl,
\end{align*}
which proves the optimality of $\Pi$ and completes the proof of Theorem \ref{thm:b01}.
\qed

\bigskip

\dowod[of Corollary \ref{cor:c01}]
First notice that $\ubl$ is the optimal value for the problem with proportional transaction costs. Indeed, if in the proof of Theorem \ref{thm:e01} we put $w_\beta(\pi_-, z) = m_\beta - \tl v_\beta(\pi_-, z)$, we obtain an analog of (\ref{eqn:e05}) with function $w$ depending on $\pi_-, z$ and $\ubl$ as above. Consequently $\ubl$ is the optimal value for the problem with proportional transaction costs and the optimal strategy for the proportional transaction costs depends only on the current state of the processes $(\pi_-(t))$ and $(Z(t))$.

Let $\Pi$ be the optimal portfolio for the case with fixed and proportional transaction costs
(as defined in Theorem \ref{thm:b01}). Denote by $\tl X_-^\Pi(t)$ the wealth of the portolio governed
by $\Pi$ when the fixed term of the transaction cost function is equal to $0$. Obviously $\tl X_-^\Pi(t) \ge X_-^\Pi(t)$ and
$$
\lim_{T \to \infty} \frac{1}{T} \ee^z \ln \tl X_-^\Pi(t) \ge \ubl.
$$
Since $\ubl$ is the optimal value for the problem with proportional transaction costs we have the opposite inequality.

Proof of (iii) follows directly from the proof of Theorem \ref{thm:b02}.
\qed

\section{Extensions}
\label{sec:g}
The paper can be extended twofolds. First consider a generalization with respect to the cost function $\tl c$. Assume that the cost function $\tl c$ is subadditive and satisfies
\begin{align}
\label{eqn:g01}
\tl c(N_1, N_2, S) &\ge \sum_{i=1}^d \Big( c^1_i S^i (N_1^i - N_2^i)^+ + c^2_i S^i (N_1^i - N_2^i)^- \big)\\
\label{eqn:g02}
\tl c(N_1, N_2, S) &\le
\sum_{i=1}^d \Big( c^1_i S^i (N_1^i - N_2^i)^+ + c^2_i S^i (N_1^i - N_2^i)^-
\big) + C
\end{align}
for some $C\ge 0$ and $c^1_i, c_2^i \in [0,1)$, $i=1, \ldots d$. If the cost function
in the right-hand side of (\ref{eqn:g02}) satisfies (A6) than there exists an optimal portfolio of the form presented in Theorem \ref{thm:b01}. Moreover, the portfolio optimal for the cost
\begin{equation}
\label{eqn:g03}
\sum_{i=1}^d \Big( c^1_i S^i (N_1^i - N_2^i)^+ + c^2_i S^i (N_1^i - N_2^i)^-
\big) + C
\end{equation}
is optimal for $\tl c$ as well. To see this let us denote by $\hat J^{\pi_-, x_-, z}(\Pi)$ the functional (\ref{eqn:a06}) for the cost function $\tl c$, by $J^{\pi_-, x_-, z}(\Pi)$ the functional (\ref{eqn:a06}) for the cost function (\ref{eqn:g03}), and finally by $\tl J^{\pi_-, z}(\Pi)$ the functional (\ref{eqn:a06}) for the cost function (this a proportional cost)
\begin{equation}
\label{eqn:g04}
\sum_{i=1}^d \Big( c^1_i S^i (N_1^i - N_2^i)^+ + c^2_i S^i (N_1^i - N_2^i)^-
\big).
\end{equation}
Easily, for any portfolio $\Pi \in \ga^z$ we have
$$
\tl J^{\pi_-, z}(\Pi) \ge \hat J^{\pi_-, x_-, z}(\Pi) \ge J^{\pi_-, x_-, z}(\Pi).
$$
This implies that
$$
\sup_{\Pi \in \ga^z} \tl J^{\pi_-,  z}(\Pi)
\ge
\sup_{\Pi \in \ga^z} \hat J^{\pi_-, x_-, z}(\Pi)
\ge
\sup_{\Pi \in \ga^z} J^{\pi_-, x_-, z}(\Pi).
$$
Since, by virtue of Theorem \ref{thm:b01} and Corollary \ref {cor:c01} there exists a constant $\lambda$ such that
$$
\lambda = \sup_{\Pi \in \ga^z} \tl J^{\pi_-, z}(\Pi)
=\sup_{\Pi \in \ga^z} J^{\pi_-, x_-, z}(\Pi)
$$
we conclude that $\lambda = \sup_{\Pi \in \ga^z} \hat J^{\pi_-, x_-, z}(\Pi)$. Moreover, due to Corollary \ref{cor:c01} the optimal portfolio for the functional $J^{\pi_-, x_-, z}$
is also optimal for $\tl J^{\pi_-, z}$. Therefore, it is also optimal for $\hat J^{\pi_-, x_-, z}$. Notice now that the cost function (\ref{eqn:a01a}) satisfies (\ref{eqn:g01}) and (\ref{eqn:g02}). Therefore, Theorem \ref{thm:b01} extends to this important case.

The results of this paper can be applied to an incomplete information case and extend \cite{Palczewski2}. Let us first sketch some motivation for this development. It is well known that investors do not have full information about variables influencing the economy. It is due to errors in statistical data or simply due to inaccesibility of some information. Therefore, it is natural to extend our model to cover the case where a number of economic factors is either observable with noise or not observable at all. For simplicity we restrict ourselves to the case when a group of factors can be precisely observed and the rest is not observable. However, our results can be extended to a more general.

Following the above remark assume that the space of economic factors $E$ is a direct sum of metric spaces $E^1$, $E^2$ with Borel $\sigma$-algebras $\bse^1$,
$\bse^2$.
Therefore, $\big(Z(t)\big)$ has a unique decomposition into $\big(Z^1(t), Z^2(t)\big)$. We shall treat $E^1$ as the observable part of the economic factor space and $\big(Z^1(t)\big)$ as the observable factor process. The process $\big(Z^2(t)\big)$ is the unobservable factor process. We denote by $\flm_t, \flz^1_t, \flz^2_t$ filtrations generated, respectively, by $\big(\zeta(t)\big)$, $\big(Z^1(t)\big)$ and $\big(Z^2(t)\big)$.
Denote by $\fly_t$ the filtration generated by $\flm_t$ and $\flz^1_t$ and
by $\tl \laa^z$ the space of $\fly_t$-adapted portfolios admissible for $z$, i.e. $\tl \laa^z \subseteq \laa^z$. Our aim is to prove existence of optimal strategy maximizing the functional
$$
J^{\pi_-, x_-, z^1, \rho}(\Pi) = \liminf_{T \to \infty} \frac{1}{T}
\ee^{z^1, \rho} \ln X_-^\Pi(T)
$$
over all strategies $\Pi \in \tl \laa$. Here $(z^1, \rho) \in E^1 \times \pe(Z^2)$ denotes the initial distribution of $\big(Z^1(t), Z^2(t)\big)$ and
$\pe(Z^2)$ stands for the space of probability measures on $(Z^2, \bse^2)$. Now, we can follow a similar reasoning as in \cite{Palczewski2} to apply Theorem \ref{thm:b01} and prove existence of an optimal portfolio. Here, however, we improve several aspects of the result; firstly, the transaction costs structure covers important examples (\ref{eqn:a01}) and (\ref{eqn:a01a}). The model setting is more general.  Moreover, in \cite{Palczewski2} the space $E^2$ has to be compact to guarantee that $\pe(E^2)$ is locally compact. Here, due to a different method of proof of Theorem \ref{thm:b01} we allow $E^2$ to be a general complete separable metric space (in this case, $\pe(E^2)$ is also a complete separable metric space). For further details see \cite{Palczewski2}.

\section{Appendix}
\dowod[of Lemma \ref{lem:b01}]
First inequality is a direct consequence of the second one. Denoting $\delta = e(\pi_-, \pi, x_-)$,
by Lemma \ref{lem:a01} $\delta \ge 0$ and  $c(\pi_-, \delta \pi) + \frac{C}{x_-} + \delta \ge 1$.
Noticing that $c(\pi_-, \delta \pi) \le d \sum_{i=1}^d |\pi_-^i - \delta \pi^i|$ we obtain
$1 \le d (1-\delta) + 2d\delta + \frac{C}{x_-} + \delta$, which easily leads to the desired inequality.
\qed

\dowod[of Lemma \ref{lem:d01}]
We shall prove (i) by contradiction: assume that $\tl e(\pi_-, \pi) < e(\pi_-, \pi, x_-)$.
Noticing
$a^+ - b^+ \le (a-b)^+$ and $a^- - b^- \le (a-b)^-$ we obtain 
$|c(\pi_-, \delta_2 \pi) - c(\pi_-, \delta_1 \pi)| \le |\delta_2 - \delta_1| \max_i (c^1_i, c^2_i)$
for
$\delta_1, \delta_2 \in [0,1]$, we have 
$$
0 \le e(\pi_-, \pi, x_-) - \tl e(\pi_-, \pi) \le \big( e(\pi_-, \pi, x_-) - \tl e(\pi_-, \pi) \big) \max_i (c^1_i, c^2_i)
- \frac{C}{x_- }.
$$
It gives the estimate $1 + \frac{C}{x_- \  \big( e(\pi_-, \pi, x_-) - \tl e(\pi_-, \pi) \big)}
\le \max_i (c^1_i, c^2_i)$, which contradicts the assumption that $c^1_i, c^2_i \in [0,1)$.
The proof of $e(\pi_-, \pi, x_-) \le e(\pi_-, \pi, \tl x_-)$ can be done in an analogous way.
Statement (ii) follows immediately from the inequality
$$
\tl e(\pi_-, \pi) - e(\pi_-, \pi, x_-) \le
\big( \tl e(\pi_-, \pi) - e(\pi_-, \pi, x_-) \big) \max_i c^1_i + \frac{C}{x_-}.
$$
For (iii) we apply the inequality $\ln (1+x) \le x$ for $x > 0$.
\qed

\dowod[of Corollary \ref{col:d01}]
For a given $\pi_- \in \es$, $z \in E$ and $\tl x_- \le  x_-$
$$
v_\beta(\pi_-, \tl x_-, z) - v_\beta(\pi_-, x_-, z) \le
\sup_{\Pi \in \ga^z} \big\{ J_\beta^{\pi_-, \tl x_-, z}(\Pi) - J_\beta^{\pi_-, x_-, z}(\Pi) \big\}.
$$
Therefore, the result follows from the observation that 
$J_\beta^{\pi_-, \tl x_-, z}(\Pi) - J_\beta^{\pi_-, x_-, z}(\Pi) \le 0$ for any $\pi \in \ga^z$.
\qed

\dowod[of Lemma \ref{lem:c02}]
Let $\underline{e} = \inf_{\pi_-, \pi \in \es} \tl e(\pi_-, \pi)$. Since $\max_i(c^1_i, c^2_i) < 1$, we have $\underline{e} > 0$. Fix $z, z' \in E$ and $\pi_-, \pi'_- \in \es$. Denote by $\Pi$ the portfolio optimal for $\tl v_\beta(\pi_-, z)$, and
by $\Pi'$ the portfolio optimal for $\tl v_\beta(\pi'_-, z')$ (they exist by Theorem \ref{thm:c01}). The corresponding proportion processes $\pi^{\Pi,z}_-(t)$,
$\pi^{\Pi', z'}_-(t)$ will be written as $\pi_-(t)$, $ \pi'_-(t)$ and the corresponding wealth processes $X^{\Pi,z}_-(t)$, $X^{\Pi', z'}_-(t)$ as $X_-(t), X'_-(t)$. We have then
\begin{align*}
\tl v_\beta(\pi_-, z) - &\tl v_\beta(\pi'_-, z') \\
&=
\sum_{t=0}^{n-1} \beta^t \ee^z h\big(\pi_-(t), z(t)\big)
+
\sum_{k=1}^\infty \ee^z \Big\{ \ind{\tau_k < n} \beta^{\tau_k} \ln \tl e\big(\pi_-(\tau_k), \pi_k\big) \Big\}\\
&-
\sum_{t=0}^{n-1} \beta^t \ee^{z'} h\big(\pi'_-(t), z'(t)\big)
-
\sum_{k=1}^\infty \ee^{z'} \Big\{ \ind{\tau_k < n} \beta^{\tau_k} \ln \tl e\big(\pi'_-(\tau_k), \pi_k\big) \Big\}\\
&+
\beta^n \Big( \ee^z \tl v_\beta\big(\pi_-(n), z(n)\big)
- \ee^{z'} \tl v_\beta\big(\pi'_-(n), z'(n)\big) \Big).
\end{align*}
There are at most $n$ transactions between $0$ and $n-1$, since by subadditivity of the cost function it is never optimal to have more than one transaction at a moment. Hence,
\begin{multline*}
\tl v_\beta(\pi_-, z) - \tl v_\beta(\pi'_-, z') \\
\le
n \norma{h}_{sp} - n \ln \underline e + \beta^n \Big( \ee^z \tl v_\beta\big(\pi_-(n), z(n)\big)
- \ee^{z'} \tl v_\beta\big(\pi'_-(n), z'(n)\big) \Big),
\end{multline*}
where $\norma{h}_sp = \sup h - \inf h$ is a span seminorm. Choose arbitrary $\pi^* \in \es$ and observe that
\begin{align*}
\ee^z \tl v_\beta\big(\pi_-(n), z(n)\big)
- \ee^{z'} \tl v_\beta\big(\pi'_-&(n), z'(n)\big)\\
&\le
\ee^z \big\{ \tl v_\beta\big(\pi_-(n), z(n)\big) - \tl v_\beta\big(\pi^*, z(n)\big) \big\}\\
&+ \ee^{z'} \big \{ \tl v_\beta\big(\pi^*, z'(n)\big) - \tl v_\beta\big(\pi'_-(n), z'(n)\big) \big\}\\
&+ \ee^{z} \tl v_\beta\big(\pi^*, z(n)\big) - \ee^{z'} \tl v_\beta\big(\pi^*, z'(n)\big).
\end{align*}
Since $\tl v_\beta(\pi_-, z) - \tl v_\beta(\pi'_-, z) \le - \ln \tl e(\pi, \pi')$, we have
\begin{align*}
\ee^z \big\{ \tl v_\beta\big(\pi_-(n), z(n)\big) - \tl v_\beta\big(\pi^*, z(n)\big) \big\}
&\le - \ln \underline e, \\
\ee^{z'} \big \{ \tl v_\beta\big(\pi^*, z'(n)\big) - \tl v_\beta\big(\pi'_-(n), z'(n)\big) \big\}
&\le - \ln \underline e.
\end{align*}
Notice that
$$
\ee^{z} \tl v_\beta\big(\pi^*, z(n)\big) - \ee^{z'} \tl v_\beta\big(\pi^*, z'(n)\big) = \int_E \tl v_\beta(\pi^*, y) \ q(dy),
$$
with $q = P^n(z, \cdot) - P^n(z', \cdot)$.
Let $\Gamma \in \bse$ be the set from the Hahn-Jordan decomposition of the signed measure $q$, i.e. $q$ is non-negative on $\Gamma$
and non-positive on $\Gamma^c$. By (A3)
$$
\int_E \tl v_\beta(\pi^*, y) \ q(dy) \le \norma{\tl v_\beta(\pi^*, \cdot)}_{sp}\ q(\Gamma)
\le \kappa \ \norma{\tl v_\beta(\pi^*, \cdot)}_{sp}.
$$
Consequently,
$$
\tl v_\beta(\pi_-, z) - \tl v_\beta(\pi'_-, z') \le
n \norma{h}_{sp} - (n + 2) \ln \underline e + \kappa \norma{\tl v_\beta(\pi^*, \cdot)}_{sp}.
$$
Since $\pi_-, \pi'_- \in \es$ and $z, z' \in E$ were arbitrary we obtain
$$
\norma{\tl v_\beta(\pi^*, \cdot)}_{sp} \le n \norma{h}_{sp} - (n + 2) \ln \underline e + \kappa \norma{\tl v_\beta(\pi^*, \cdot)}_{sp},
$$
which yields the desired result.
\qed

\begin{mythebibliography}{8}
\bibitem{aase} Aase K, \O{}ksendal B (1988) {Admissible investment strategies in continuous trading}, Stoch. Proc. Appl. 30:291-301
\bibitem{Akian} Akian M, Sulem A, Taksar M (2001) {Dynamic optimization of long term growth rate for a portoflio with transaction costs -- the logarithmic utility  case.} Math. Finance 11.2: 153 - 188
\bibitem{algoet} Algoet PH, Cover TM (1988) {Asymptotic optimality and asymptotic equipartition properties of log-optimum investment.} Ann. Prob. 16:876-898
\bibitem{Arapost} Arapostathis A et al. (1993) {Discrete-time controlled Markov processes with average cost criterion: a survey.} SIAM J. Control Optim, 31.2: 282 - 344
\bibitem{Bielecki2} Bielecki TR, Pliska SR (1999) {Risk Sensitive Dynamic Asset Management.}  Appl. Math. Optim. 37: 337 - 360
\bibitem{Bielecki1} Bielecki TR, Pliska SR, Sherris M (2004)
{Risk sensitive asset allocation.} J. Econ. Dyn. Control 24: 1145-1177
\bibitem{brown} Brown LD, Purves R (1973) {Measurable Selections of Extrema} Ann. Stat. 1.5: 902-912
\bibitem{donsker} Donsker MD, Varadhan SRS (1976) {Asymptotic Evaluation of Certain Markov Process Expectations for Large Time - III.} Comm. Pure Appl. Math. 29: 389-461
\bibitem{Doob} Doob JL (1953)  {Stochastic Processes.} Wiley
\bibitem{duffie} Duffie D (2001) {Dynamic Asset Pricing Theory.} Princeton University Press
\bibitem{duncan} Duncan T, Pasin-Duncan B, Stettner \L (2000) {Adaptive control of discrete time Markov processes by large deviations method.} Applicationes Mathematicae 27.3: 265-285
\bibitem{fleming} Fleming WH, Sheu SJ (2000) {Risk-sesitive control and an optimal investment model.}, Math. Finance 10.2: 197-213
\bibitem{gerenc} Gerencs\'{e}r L, R\'{a}sonyi M, V\'{a}g\'{o} Zs (2005) {Log-optimal currency portoflios and control Lyapunov exponents.} 44th IEEE Conference on Decision and Control and European Control Confrence ECC 2005: 1746-1769
\bibitem{gonzalez} Gonzalez-Trejo JI, Hernandez-Lerma O, Hoyos-Reyes LF (2003) {Minimax control of discrete-time stochastic systems.} SIAM J. Control Optim, 41.5: 1626-1659
\bibitem{Hernandez} Hernandez-Lerma O, Lasserre JB (1996) {Discrete-Time Markov Control Processes.} Springer
\bibitem{Hernandez1} Hernandez-Lerma O, Lasserre JB (1999) {Further Topics on Discrete-Time Markov Control Processes.} Springer
\bibitem{inoue} Inoue A, Nakano Y (2005) {Optimal long term investment model with memory.} to appear in Appl. Math. Optim.
\bibitem{Iyengar} Iyengar G (2005) {Universal investment in markets with transaction costs.}
Math. Finance 15.2: 359-371
\bibitem{jaskiewicz} Ja\'skiewicz A, Nowak AS (2006) {On the optimality equation for the average cost Markov control processes with Feller transition probabilities.} J. Math. Anal. Appl. 316: 495-509
\bibitem{jaskiewicz2} Ja\'skiewicz A, Nowak AS (2006) {Zero-sum ergodic stochastic games with Feller transition probabilities.} SIAM J. Control Optim. 45.3: 773-789
\bibitem{kelly} Kelly JL (1956) {A New Interpretation of Information Rate} Bell System Technical Journal 35: 917-926 
\bibitem{kuroda} Kurod K, Nagai H (2002) {Risk-sensitive portfolio optimization on infinite time horizon.} Stoch. Stoch. Rep. 73: 309-331
\bibitem{liptser} Liptser R (1996) {Large Deviations For Occupation Measures Of Markov Proceses: Discrete Time, Noncompact Case.} Th. Prob. Appl. 41.1: 35-54
\bibitem{luenberger} Luenberger DG (1998) {Investment science.} Oxford University Press
\bibitem{Palczewski2} Palczewski J, Stettner \L (2007) {Maximization of the portfolio growth rate under fixed and proportional transaction costs.} Communications in Information and Systems 7.1: 31-58
\bibitem{Palczewski1} Palczewski J, Stettner \L (2007) {Impulsive control of portfolios.} Appl. Math. Optim. 56.1: 67-103
\bibitem{Platen} Platen E (2006) {A Benchmark Approach to Finance.} Math. Finance 16: 131-151
\bibitem{Serfozo} Serfozo R (1982) {Convergence of Lebesgue integrals with varying measures.}
Sankhya Ser. A 44.3: 380 - 402
\bibitem{Schal} Sch\"al M (1993) {Average optimality in dynamic programming with general state space.} Math. Oper. Res. 18.1: 163 - 172
\bibitem{Stettner} Stettner \L (1983) {On impulsive control with long run average cost criterion.} Studia Mathematica 76.3: 279 - 298
\bibitem{Stettner2} Stettner \L (2005) {Discrete Time Risk Sensitive Portfolio Optimization with Consumption and Proportional Transaction Costs.} Applicationes Mathematicae 32.4: 395 - 404
\bibitem{thorpe} Thorp EO (1975) {Portfolio choice and the Kelly criterion.} in: Stochastic Optimization Models in Finance, Ziemba WT and Wickson RG, eds., Academic Press, New York: 599-619
\end{mythebibliography}

\end{document}